\documentclass[aps,prl,twocolumn,groupedaddress,longbibliography]{revtex4-1}

\usepackage{color, graphicx}
\usepackage{hyperref}
\usepackage{amsmath, amssymb, amsfonts}
\usepackage{mathtools}

\newcommand{\cL}{\mathcal{L}}
\newcommand{\cW}{\mathcal{W}}
\newcommand{\cH}{\mathcal{H}}
\newcommand{\cA}{\mathcal{A}}
\newcommand{\cB}{\mathcal{B}}
\newcommand{\cM}{\mathcal{M}}
\newcommand{\bR}{\mathbf{R}}
\newcommand{\br}{\mathbf{r}}
\newcommand{\bx}{\mathbf{x}}
\newcommand{\by}{\mathbf{y}}
\newcommand{\bv}{\mathbf{v}}

\newcommand{\bk}{\mathbf{k}}
\newcommand{\bQ}{\mathbf{Q}}
\newcommand{\bq}{\mathbf{q}}
\newcommand{\td}{2\mathrm{D}}
\newcommand{\eff}{\mathrm{eff}}
\newcommand{\intac}{\mathrm{int}}
\newcommand{\KG}{\mathrm{KG}}
\newcommand{\Pl}{\mathrm{Pl}}
\newcommand{\rB}{\mathrm{B}}
\newcommand{\ptl}{\partial}
\newcommand{\hta}{\hat{\tilde{a}}}
\newcommand{\ha}{\hat{a}}

\newcommand{\htphi}{\hat{\tilde{\phi}}}
\newcommand{\tphi}{\tilde{\phi}}
\newcommand{\hphi}{\hat{\phi}}
\newcommand{\hpsi}{\hat{\psi}}

\newcommand{\hrho}{\hat{\rho}}

\newcommand{\bpsi}{\bar{\psi}}
\usepackage{color}

\newcommand{\ket}[1]{\left|#1\right>}

\begin{document}


\title{Probing the 
scale invariance of the inflationary power spectrum 
in expanding quasi-two-dimensional 
dipolar 
condensates}


\author{Seok-Yeong Ch\"a}
\author{Uwe R. Fischer}
\affiliation{Seoul National University, Department of Physics and Astronomy, Center
for Theoretical Physics, 
Seoul 08826, Korea}


\date{\today}

\begin{abstract}
We consider an analogue de Sitter cosmos in an expanding quasi-two-dimensional Bose-Einstein condensate with dominant dipole-dipole interactions between the atoms or molecules in the ultracold gas. 
It is demonstrated that a hallmark signature of inflationary cosmology, 
the scale invariance of the power spectrum of inflaton field correlations,  
experiences strong modifications 
when, at the initial stage of expansion,  
the excitation spectrum displays a roton minimum.
Dipolar quantum gases thus furnish a viable laboratory tool to experimentally
investigate, with well-defined and controllable 
initial conditions, whether primordial oscillation spectra 
deviating from Lorentz invariance at trans-Planckian momenta 
violate standard predictions of inflationary cosmology. 


\end{abstract}

\pacs{}

\maketitle



The hypothesis of a rapid initial 
expansion of the cosmos in the inflationary scenario \cite{Guth,Linde,mukhanov2005physical} 
resolved many vexing cosmological questions plaguing other theories, 
such as the observed flatness and homogeneity of the universe, 
as well as the nonexistence of monopoles. 
However the resolution of these issues 
comes at the price of creating another potential problem  \cite{PhysRevD.63.123501}: 
Generally the period of inflation lasts so long that, at the beginning of the inflationary period, the physical
wavelengths of comoving scales which correspond to the present large-scale structure of the universe were
smaller than the Planck length. 
Thus necessarily trans-Planckian energies become involved, for which the 
physics is at present 
speculative. 
Similar issues regarding 
kinematical phenomena for quantum fields propagating on a fixed curved spacetime  
arise when tracing back Hawking radiation emission 
all the way down to the black hole horizon 
\cite{Unruh95,Corley1,Universality,Leonhardt}. 

The analogue gravity program \cite{PhysRevLett.46.1351,MattCQG,Grisha,BLV}
has been successfully theoretically implemented  in ultracold matter
for various cosmological phenomena, e.g.,  
inflaton quantum fluctuations \cite{Schuetzhold,1367-2630-7-1-248}, 
the Gibbons-Hawking effect \cite{Fedichev}, cosmological particle production \cite{BLV2003PRA,CPP,0264-9381-26-6-065012}, the cosmological constant problem \cite{Grisha,FinazziLambda,GrishaWeyl}, 
or false vacuum decay \cite{Fialko}. Importantly, recent experimental advances have allowed for  groundbreaking observations of analogues of cosmological particle production, Sakharov oscillations, black hole lasers,  
and Hawking radiation 
\cite{Jaskula,Chin,Steinhauer,Boiron,Steinhauer16}. 
In the near future, these experiments hold promise 
to realize experimental cosmology: A quantum simulation of inflation 
with reproducible initial conditions distinct from the current purely observational cosmology
of a pregiven state of the universe.
Furthermore, a major original motivation of analogue gravity, so far not experimentally investigated,  
is to probe consequences of trans-Planckian physics in a microscopically well understood setup in a regime inaccessible for quantum fields in the presence of strong real (Einsteinian or other) gravity. 
We here propose to realize this aim with dipolar Bose-Einstein condensates (BECs), 
addressing the trans-Planckian problem of inflationary cosmology. 

Going beyond contact interactions (in field theory language $\phi^4$),   
magnetic dipole-dipole interaction (DDI) dominated condensates with 
chromium \cite{Lahaye}, dysprosium \cite{Lu}, and erbium \cite{Aikawa} atoms  
have been created, and 
the future realization of electrically dipolar BECs \cite{Julienne} 
will offer even greater flexibility
in controlling the ratio of dipolar and contact interactions.
The excitation spectrum of DDI-dominated BECs displays a roton minimum \cite{PhysRevLett.90.250403,Fischer,PhysRevLett.98.030406}, and roton-induced 
dynamical effects are being 
experimentally investigated 
\cite{Pfau,Droplets}.
In addition, the significant progress in probing correlation functions to increasing accuracy \cite{Chin,Hodgman} 
pave the way for an 
exploration of the intricate many-body correlations due to the DDI. 

For certain classes of inflaton dispersion relations, displaying  deviations from Lorentz invariance at trans-Planckian scales, the  predictions of inflation, in particular the scale invariance of the power spectrum 
(SIPS) of inflaton field correlations, remain robust, while for others, they change significantly, 
cf., e.g., \cite{Niemeyer,Starobinsky2001,doi:10.1142/S0217732301004170,doi:10.1142/S0217732301004170,0264-9381-30-11-113001,PhysRevD.89.043507}.  
We will show that dipolar BECs, possessing 
trans-Planckian spectra leading to strong departures from Lorentz invariance, 
yield significant changes of the standard inflationary prediction of 
SIPS. 
To the best of our knowledge, this represents the first example within analogue gravity 
where violations of SIPS can become experimentally manifest.
 


We start with the Lagrangian density of a Bose gas comprising atoms or molecules of mass $m$,  
\begin{align}
\cL & = \frac{i\hbar}{2}\bigl( \Psi^*\partial_t{\Psi} - \partial_t{\Psi}^*\Psi\bigr) - \frac{\hbar^2}{2m} |\nabla\Psi|^2 - V_{\mathrm{ext}}|\Psi|^2  \notag \\
    & \quad - \frac{1}{2}|\Psi|^2 \int d^3\bR'\, V_{\intac}(\bR - \bR') |\Psi(\bR')|^2, \label{eqn:20150924_1}
\end{align}
where $\bR = (\br,z)$ are spatial 3D coordinates.  
The trapping potential is
$
V_{\mathrm{ext}}(\bR,t) = m\omega^2 \br^2/2 + m\omega_z^2 z^2 /2,
$
where $\omega$ and $\omega_z$ 
can be chosen functions of time. The gas is strongly confined in $z$-direction, with aspect ratio $\kappa = \omega_z/\omega \gg 1$ over the whole time evolution.
The 
interaction reads 
$
V_{\intac}(\bR - \bR') = g_c \delta^{(3)}(\bR - \bR') + V_{\mathrm{dd}}(\bR - \bR'),
$
where $g_c$ 
is the contact interaction coupling,  
and $V_{\mathrm{dd}}(\bR) = 3g_d[(1 - 3z^2/|\bR|^2)/|\bR|^3]/4\pi$ for dipoles polarized perpendicular to the plane. 
{We have a {\em scaling law}  
$V_{\rm int} (\Lambda \bR) = \Lambda^\alpha V_{\rm int} (\bR)$
 \cite{1367-2630-12-11-113005} for a combined 3D contact and dipolar potential with $\alpha=-3$. 
Note that the scaling equation \eqref{eqn:20150925_3} below is thus 3D. 
To ensure stability in the DDI dominated regime \cite{Fischer}, we impose the system to stay 
sufficiently close to the quasi-2D regime during expansion.}
We integrate out the $z$ dependence by assuming a Gaussian, 
$\rho_z(z) = (\pi d_z^2)^{-1/2}  
\exp\left[-{z^2}/{d_z^2}\right]$, where $d_z = b(t) d_{z,0}$, and $b(t)$ is the scale factor in Eq.\,\eqref{scalingEqs} \cite{renorm}, where 
$d_{z,0} = \sqrt{\hbar/m\omega_{z,0}}$.
The reduced contact coupling and lower-dimensional interaction are, respectively, then given by
$g_c^{\td} = g_c/\sqrt{2\pi}d_z$, $V_{\rm int}^{\td}(\br - \br') = \int dzdz'\, V_{\rm int}(\bR - \bR')\rho_z(z)\rho_z(z')$ \cite{suppl}. 

Employing the conventional scaling transformation to describe the evolution of the BEC 
upon changing the trapping or the coupling constants \cite{PhysRevA.54.R1753,CastinDum,1367-2630-12-11-113005},
\begin{align}
\bx & := \frac{\br}{b(t)},\;\;\quad \tau := \int_0^t \frac{ dt'}{b^2(t')}, \notag \\
\Psi(\br,t) & := e^{i(mr^2/2\hbar)(\ptl_t b/b)}  
\frac{\psi(\bx,\tau)}{b}, 
\label{scalingEqs}
\end{align}
we obtain the nonlocal Gross-Pitaevski\v\i ~equation 
\begin{align}
i\hbar \ptl_\tau \psi & = -\frac{\hbar^2}{2m} \nabla_x^2 \psi + f^2\biggl(\frac{m}{2}\omega_0^2 x^2 + g_{c,0}^{\td} |\psi|^2 \notag\\
                      & \quad + \int d^2\bx'\, V_{{\rm dd},0}^{\td}(\bx - \bx')|\psi(\bx')|^2\biggr)\psi. \label{eqn:20150925_5}
\end{align}
We combined all remaining time dependences into a single factor $f=f(t)$, imposing 
\cite{PhysRevA.79.033601} [$b(0) = f(0) = 1$]
\begin{equation} \label{eqn:20150925_3}
f^2= \frac{ b^3 \partial^2_t b+  b^4 \omega^2(t)}{\omega_0^2} = \frac{g_c(t)}{g_{c,0}b} = \frac{g_d(t)}{g_{d,0}b}.
\end{equation}
To separate off the effective contact interaction contribution $g_0^{\eff}$, 
we put $V_{\intac,0}^{\td}(\bx - \bx') = g_{c,0}^{\td}\delta^{(2)}(\bx-\bx') + V_{{\rm dd},0}^{\td}(\bx - \bx') := 
g_0^{\eff}\delta^{(2)}(\bx-\bx') + U_0^{\td}(\bx-\bx')$ 
\cite{suppl}.



The planar cloud size greatly exceeds the wavelengths of relevant Bogoliubov excitations in the plane.
Especially near the center of cloud, density gradients are thus negligible,  
 and we approximate the 2D {comoving} density $\rho_0\simeq $\,const.
The  Bogoliubov equations for density and phase fluctuations read, with comoving momentum $\bk$,    
\begin{eqnarray} 
(\ptl_\tau + i \bv_{\rm com} \cdot \bk) \delta \rho_\bk & = & \frac{\hbar \rho_0}{m} k^2 \delta \phi_\bk, \nonumber\\
(\ptl_\tau + i \bv_{\rm com} \cdot \bk) \delta \phi_\bk & = & -\frac{f^2g_0^{\eff}}{\hbar} \cW_k \delta \rho_\bk,
\label{eqn:20151102_1}\\
\cW_k = \frac{\zeta^2}{4Af^2} + \frac{1}{g_0^{\eff}}{V}_{\intac,0}^{\td} (\zeta) , & & 
 A = \frac{m c_0^2}{\hbar\omega_{z,0}} 
 \label{eqn:20151102_2}, 
\end{eqnarray} 
and $c_0 = \sqrt{{g_0^{\eff} \rho_0}/{m}}$. 
Furthermore, ${V}_{\intac,0}^{\td}(\zeta)$ is the Fourier transform of $V_{\intac,0}^{\td}$, with dimensionless $\zeta = k d_{z,0}$, and $\bv_{\rm com}$ is the comoving frame velocity \cite{suppl}; 
we assume it to be negligibly small in what follows, $\bv_{\rm com} \simeq 0$. 

Solving the second equation in \eqref{eqn:20151102_1} for $\delta \rho_{\bk}$ and substituting into the first equation yields
\begin{equation} \label{eqn:20151102_6}
\delta \ddot{\tilde{\phi}}_k + \biggl( 2\frac{\dot{a}}{a} - \frac{\dot{\cW}_k}{\cW_k}\biggr)\delta \dot{\tilde{\phi}}_k + \biggl( \frac{c_0 k}{a}\biggr)^2 \cW_k \delta \tilde{\phi}_k = 0,
\end{equation}
where overdot denotes $\tau$ derivatives, and we rescaled $\delta \tilde{\phi} = \Omega^{-1/2} \delta \phi,\; \Omega = c_0^2 m^2/\hbar^2 \rho_0$ for later convenience. We introduce 
the Friedmann-Robertson-Walker (FRW) cosmological scale factor by $a(t) := 1/f(t)$,  see  below for a detailed discussion.

\begin{figure}[bt]
\centering
\includegraphics[width=0.42\textwidth]{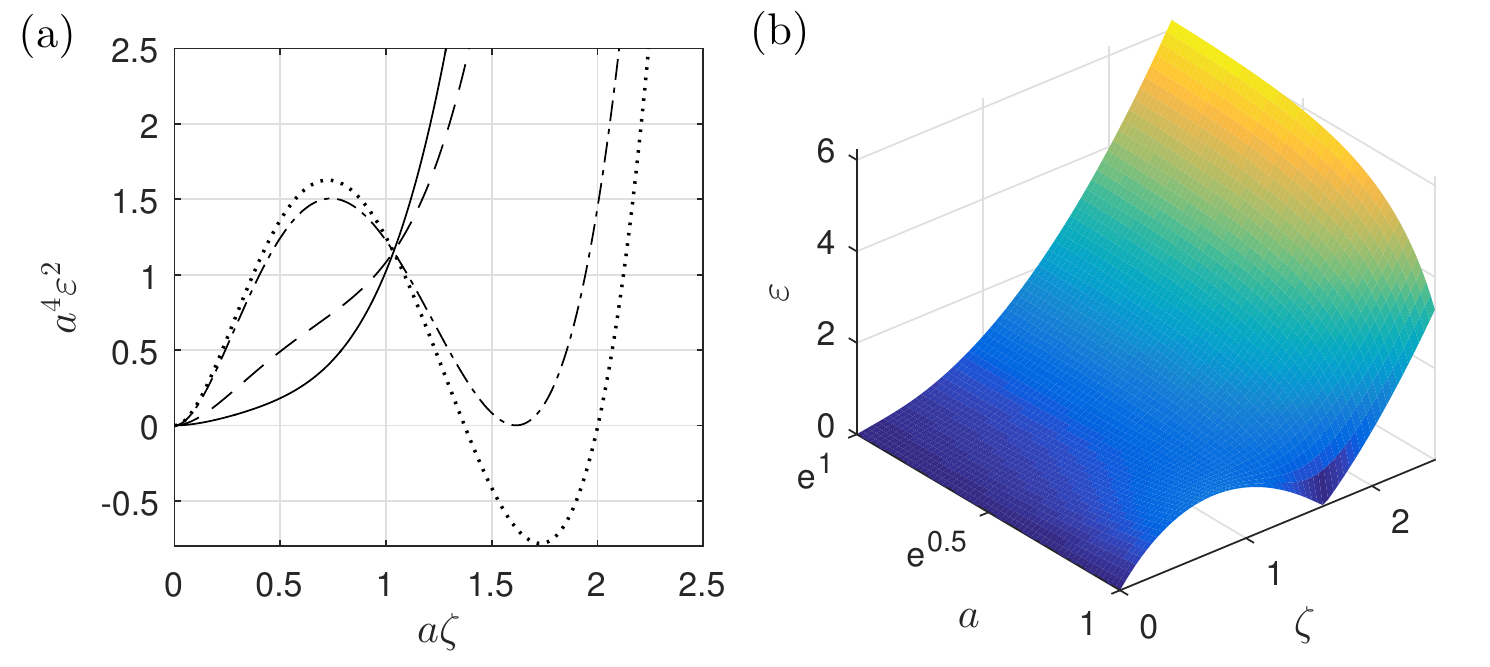}
\caption{(a) Squared Bogoliubov excitation energy in units of $\hbar^4/4m^2d_{z,0}^4$, 
for DDI domination, $R = \sqrt{\pi/2}$.
Counting from bottom to top at small $\zeta$, $A$ in \eqref{eqn:20151102_2} is $A_c/10,\; A_{\mathrm{min}}, \; A_c$, and $1.1A_c$.
For $A > A_{\mathrm{min}} = 1.249$, the spectrum develops a roton minimum, and becomes unstable for $A>A_c = 3.4454$.
(b) Time evolution of the Bogoliubov spectrum 
in the course of expansion at criticality  
$A = A_c$.
Initially, a roton minimum occurs, disappearing at late times.}
\label{fig:Bogol_Spec}
\end{figure}

In an adiabatic regime 
\cite{PhysRevA.79.033601}, 
momentarily ignoring time derivatives of $a$ and $\cW_k$, the Bogoliubov dispersion is 
\begin{equation} \label{eqn:20150925_13}
\frac{\varepsilon^2}{(\hbar \omega_{z,0})^2} 
=  
\frac{A \zeta^2}{a^2} \left( 1 - \frac{3R}{2}\zeta w \left[ \frac{\zeta}{\sqrt{2}}\right] \right)
+\frac{\zeta^4}{4},
\end{equation}
where the $w$ function $w(z)=\exp[z^2](1-{\rm erf}[z])$ contains the error function 
\cite{Fischer,abr65}. 
The dimensionless parameter,
$R = g_{d,0}/(d_{z,0}g_0^{\eff}) = {\sqrt{\pi/2}}/({1 + g_{c,0}/2g_{d,0}})$, 
ranges from $R = 0$ if $g_{d,0}/g_{c,0} \to 0$ to $R = \sqrt{\pi/2}$ for $g_{d,0}/g_{c,0} \to \infty$. 
In the latter DDI dominated case, the spectrum displays a ``roton'' minimum, which touches zero when $A$ is equal to the critical value $A_c = 3.4454$, 
see Fig.\ref{fig:Bogol_Spec}.

In the long-wavelength limit, $\cW_k\to 1$, and \eqref{eqn:20151102_6} becomes
\begin{equation} \label{eqn:20150926_6}
\delta\ddot{\tilde{\phi}}_k + 2\frac{\dot{a}}{a}\delta \dot{\tilde{\phi}}_k + \biggl( \frac{c_0k}{a}\biggr)^2\delta \tilde{\phi}_k = 0.
\end{equation}
The gravitational analogy can now be established by introducing the analogue space-time  line element 
\cite{BLV,doi:10.1142/S0218271803004092,CPP} 
\begin{equation} \label{eqn:20150925_20}
ds^2 =  \Omega^{-2} (c_0^2 d\tau^2 - a^2 d\bx^2 ).
\end{equation}
Then, \eqref{eqn:20150926_6} becomes the wave equation for a massless, minimally coupled free scalar field $\Box\delta \tilde{\phi} = (1/\sqrt{|\tilde{g}|})\ptl_\mu(\sqrt{|\tilde{g}|}\tilde{g}^{\mu\nu} \ptl_\nu \delta \tilde{\phi}) = 0$, in the (2+1)D FRW spacetime of Eq.\,\eqref{eqn:20150925_20} \cite{suppl}. 

{The simple possibility of $g_c(t)\propto g_d(t)\propto \exp[-2Ht]$ 
to realize an effective de Sitter (dark energy dominated) cosmos, $a = \exp[Ht]$, while  
having the advantage that the gas does not need to expand [$b(t)=1\,\forall\, t$ and thus $\tau :=t$], comes with the experimental difficulty that 
{both} couplings need to vary exponentially rapidly in lab time,   
see \,Eqs.\,\eqref{scalingEqs} and \eqref{eqn:20150925_3}. 
While this is, in principle, possible \cite{PhysRevLett.89.130401}, also cf.\,Ref.\,\cite{Natu}, 
we keep for the below discussion $g_c$ as well as $g_d$ 
constant; then $a^2(t) =b(t)$. 
For de Sitter expansion, $a(\tau) = 1/f(\tau) = e^{H\tau}$, and thence in the lab,} 
\begin{equation} \label{eqn:20150926_5}
b(t) = \sqrt{4Ht + 1}, \quad \omega^2(t) = {\omega_0^2}/{b^5} + 4H^2/b^4 .
\end{equation}
The radial condensate velocity then scales  
as $v=2H r/b^2$ and the kinetic energy per particle, relative to $\omega_{z,0}$,  
as $A/b^2$. It thus decreases 
$\propto \omega_z$, ensuring proximity to the quasi-2D limit $\forall\,t$. The much slower (in comoving $\tau$ space) 
pre-de Sitter stage of cosmic expansion, $t<0$, is conceived such that it $\approx$ adiabatically  
leads to $\partial_t b(0) =2H$, 
and can 
be used to simulate as well the radiation- [$a(\tau)\propto \tau^{1/2}$] and matter-dominated 
[$a(\tau)\propto \tau^{2/3}$] eras 
\cite{peacock1999cosmological},  
by appropriately 
tuning $\omega(t)$ and/or $g_{c,d}(t)$. 

It is noteworthy that with the 
(asymptotically square-root) expansion \eqref{eqn:20150926_5}, 
Eq.\,\eqref{eqn:20150926_6} yields an analytical solution
\begin{equation} \label{eqn:20150928_10}
\delta\tilde{\phi}_k(s) = s\sqrt{\frac{\pi\hbar VH}{4mc_0^2k^2}}\Bigl[ J_1(s) + iY_1(s) \Bigr] := h_k(s),
\end{equation}
where the variable $s = (c_0/H)/(a/k)$ measures the ratio of Hubble radius to the 
cosmic expansion-rescaled wavelength, and $J_1,Y_1$ are Bessel functions \cite{suppl}; 
$s$ starts from $\infty$ and approaches zero when $\tau$ runs 
from $-\infty$ to $\infty$, that is when conformal time $\eta := \int_\infty^\tau d\tau' c_0/a(\tau') $, 
for which $ds^2 = a^2 [ d\eta^2 - d\bx^2]$,  ranges from $-\infty$ to 0.
The {instantaneous} vacuum corresponding to the basis $h_k(s)$ is the Bunch-Davies vacuum \cite{bunch1978quantum}, 
yielding an asymptotic Minkowski vacuum in the (formally) infinite past  
{equivalent to the lab's  quasiparticle vacuum.} 
{It is assumed 
that the initial Bunch-Davies vacuum $\ket{0}$ 
at $\eta = -c_0/H$ ($t=\tau=0$) is during the pre-de Sitter stage 
smoothly connected to this asymptotic vacuum.
We emphasize 
that ``cosmological" quasiparticles are 
measurable: 
In \cite{suppl},  
we  establish 
the equivalence 
of representations using cosmological comoving or scaling and 
lab frame Bogoliubov quasiparticles, also cf.\,\cite{PhysRevA.79.043616, PhysRevD.82.044042}, and elaborate on the measurement process when the expansion is stopped.}

\begin{figure}[hbt]
\centering
\includegraphics[width=0.42\textwidth]{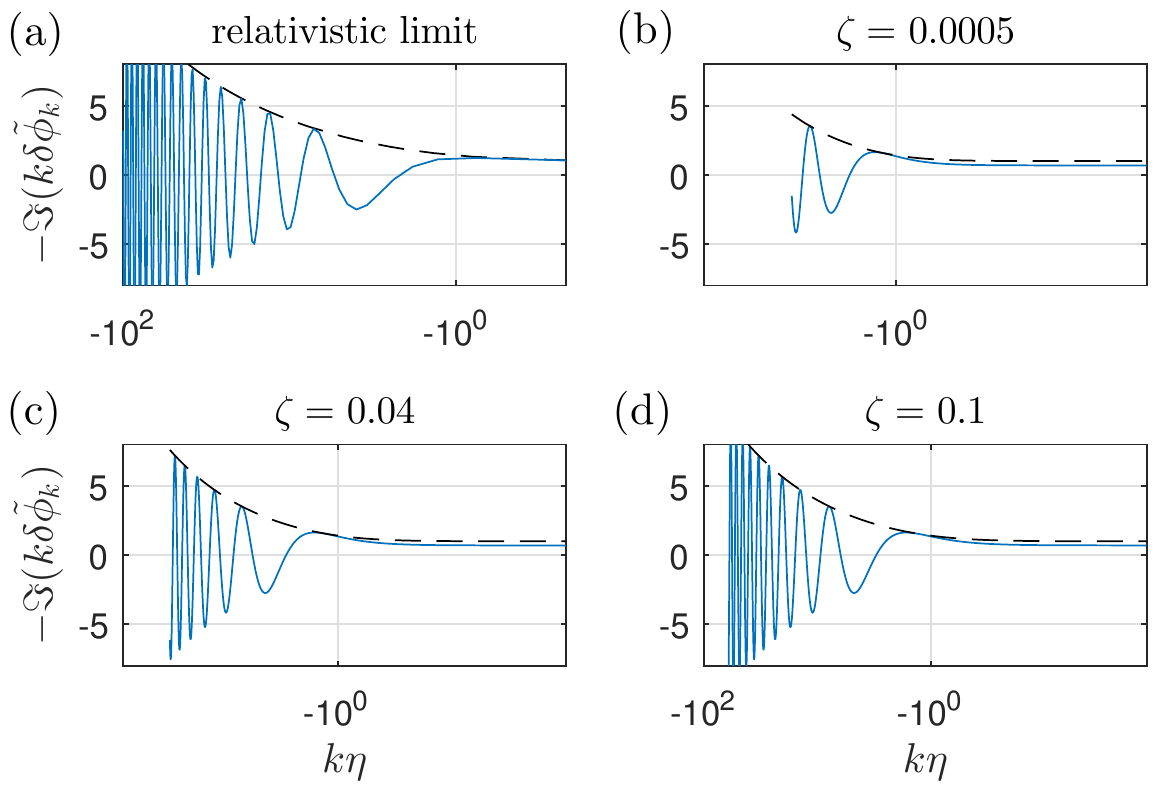}
\caption{(a) Freezing process of the inflaton mode function $kh_k$ in units of $\sqrt{\hbar HV/\pi mc_0^2} = \sqrt{HV/\pi A\omega_{z,0}}$ in terms of the wavenumber dependent logarithmic conformal time $k\eta$. Blue solid line represents the imaginary part of $kh_k$ and black dashed line represents the absolute value of $kh_k$. (a) Lorentz-invariant relativistic regime. (b)$\sim$(d) demonstrate that when the trans-Planckian spectrum is taken into account, solving Eq.\eqref{eqn:20160411_5}, 
freezing still occurs ($A = A_c/10$ and $R = \sqrt{\pi/2}$).}
\label{Fig2}
\end{figure}

The modes oscillate almost freely for $\eta \to -\infty$. 
At $k\eta = -1$ and horizon crossing, the mode freezes, 
leading to the standard theory of inhomogeneity or galaxy formation during inflation \cite{mukhanov2005physical}. 
At late times, $s\rightarrow 0$, 
and the modes do effectively not evolve anymore.
Fig.\,\ref{Fig2} shows the evolution of 
$kh_k$ as a function of $k\eta$; (b)$\sim$(d) illustrate the fact that when  trans-Planckian defomation of the spectrum is included (see below), horizon crossing and mode freezing nontrivially still occur. 

We define the 
power spectrum $P(k)$ as the Fourier transform of the correlation function \cite{peacock1999cosmological}, $\xi(\bx - \by) = \langle 0| \delta \htphi(\bx,\tau)\delta \htphi(\by,\tau)| 0 \rangle := \sum_\bk P(k) e^{i\bk\cdot(\bx - \by)}/V$, from which we have $P(k) = \langle \delta \htphi_\bk \delta \htphi_\bk^\dag \rangle/V = |h_k|^2/V$. 
At late times, $\eta \to \infty$, the power spectrum $P(k)$ converges to $\hbar H/\pi m c_0^2 k^2$ and we see that the quantity, 
\begin{equation}
\Delta^2(k) := k^2P(k),
\end{equation} 
becomes independent of $k$.
We thus obtain, after freezing, a spectrum in which $\Delta^2(k)$, the variance per $\ln k$ \cite{peacock1999cosmological,suppl}, 
is constant, 
conventionally referred to as the 
cosmological SIPS \cite{PhysRevD.1.2726, Zeldovich01101972, Peebles:1970ag}. 
Note that SIPS is not {\em per se} related to the scaling approach to describe expansion of the gas.

\begin{figure}[bt]
\centering
\hspace*{-0.5em}\includegraphics[width=0.49\textwidth]{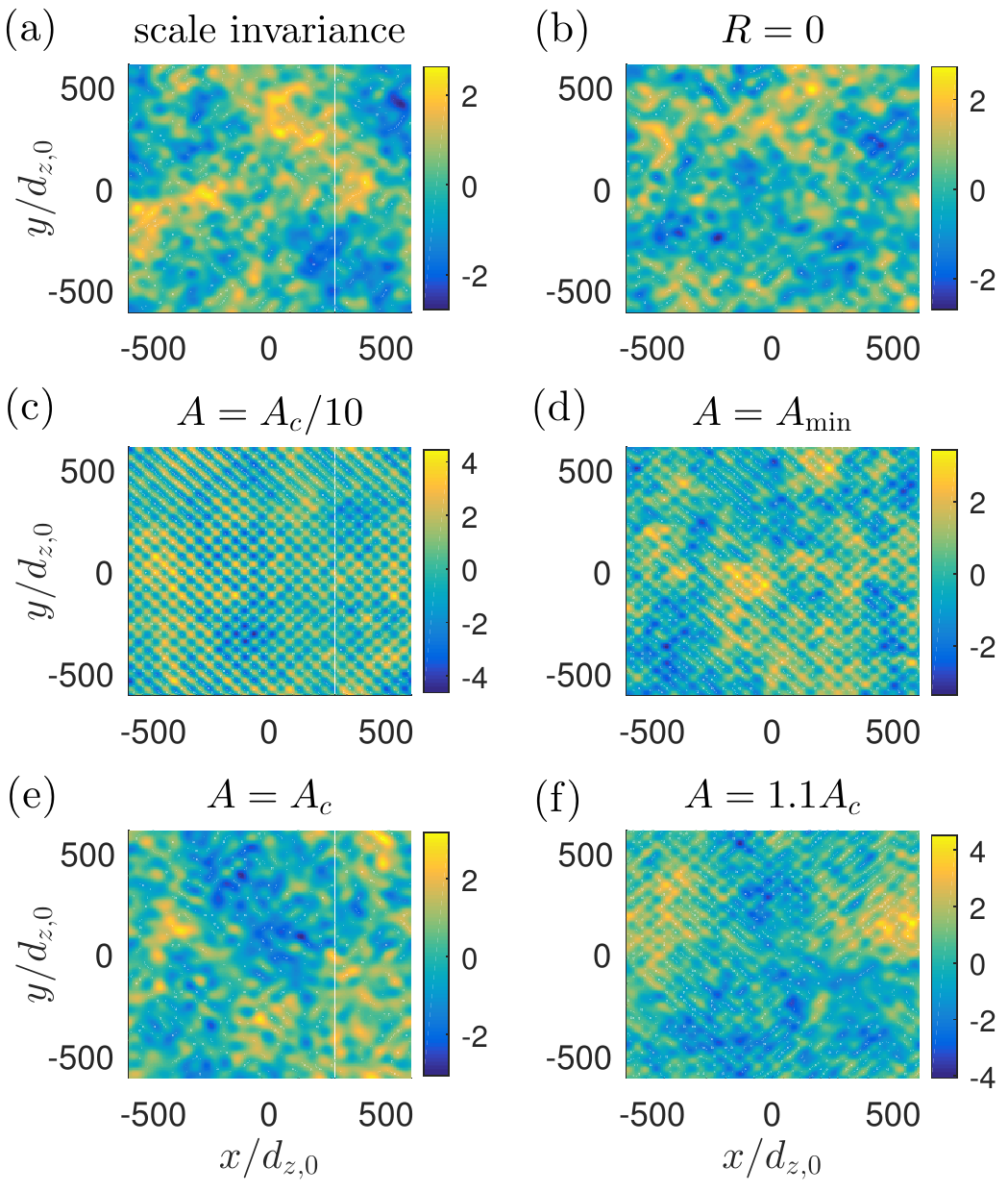}
\caption{(a) Coordinate space representation of the (real) 
field $\delta \tilde{\phi}(\bx,\tau)$,  
in units of $\sqrt{\hbar HV/\pi m c_0^2d_{z,0}^2}$ after the completion of the freezing process,
where the 2D volume of the system $V = (2\kappa_0 d_{z,0})^2$, with initial aspect ratio $\kappa_0$, and
the wavevector separation is chosen to be $\Delta k = 2\pi/2\kappa_0 d_{z,0}$.
The statistical self-similarity reveals itself by the same degree of ``wrinkliness'' on each scale.
(b) The field obtained from numerical implementation of the full Bogoliubov equations  
($A = A_c/10$, $R = 0$).
Plots (c) through (f) are for increasing $A$ and dominating DDI ($R = \sqrt{\pi/2}$).
}
\label{fig:GRF_realization}
\end{figure}

{Regarding $\delta \htphi(\bx,\tau)$ as a homogeneous and isotropic Gaussian random field \cite{mukhanov2005physical}, 
and identifying its variance by 
$\sigma_k^2 := P(k)$ \cite{suppl}, 
we obtain a real space realization 
of the scale-invariant fully relativistic limit, see 
Fig.\,\ref{fig:GRF_realization}\,(a).}  


The solution \eqref{eqn:20150928_10} represents phonons residing in the low-momentum  corner of the Bogoliubov dispersion relation [see Fig.\,\ref{fig:Bogol_Spec}\,(a)].
In order to incorporate trans-Planckian dispersion and to describe its 
influence on the small $k$ regime, we consider the more general Bogoliubov equations \eqref{eqn:20151102_6}.
We start by rewriting \eqref{eqn:20151102_6} in terms of $s$ 
\begin{equation} \label{eqn:20151216_1}
\delta \tphi''_k - \frac{G(\zeta)^2 - \frac{(a\zeta)^4}{4A}}{G(\zeta)^2 + \frac{(a\zeta)^4}{4A}}\frac{1}{s}\delta\tphi'_k + \left[ \frac{(a\zeta)^4}{4A} + G(\zeta)^2 \right] \delta \tphi_k =0,
\end{equation}
where prime denotes $s$ derivatives, and $G(\zeta)^2 := {V_{\intac,0}^{\td}}(\zeta)/{g_0^{\eff}}$.
Taking into account \eqref{eqn:20150925_13} and \eqref{eqn:20151102_2}, the linear dispersion occurs for wavenumbers satisfying
\begin{equation} \label{eqn:20160411_2}
\frac{(a\zeta)^2}{4A} \ll G(\zeta)^2\quad \mbox{($\Rightarrow$ analogue Planck scale $\zeta_{\rm PL}$}).
\end{equation}
For small $\zeta$, $G(\zeta)^2\rightarrow 1$, and Eq.\,\eqref{eqn:20160411_2} defines 
$\zeta_{\rm PL}$. 
Experiments will generally probe sub-Planckian  $\zeta$ 
that satisfy \eqref{eqn:20160411_2}. 
Therefore, we consider \eqref{eqn:20151216_1} given 
\eqref{eqn:20160411_2} is fulfilled,
\begin{equation} \label{eqn:20160411_5}
\delta \tphi''_k - \frac{1}{s} \delta \tphi'_k + G(\zeta)^2 \delta \tphi_k = 0.
\end{equation}

\begin{figure}[t]
\centering
\includegraphics[width=0.42\textwidth]{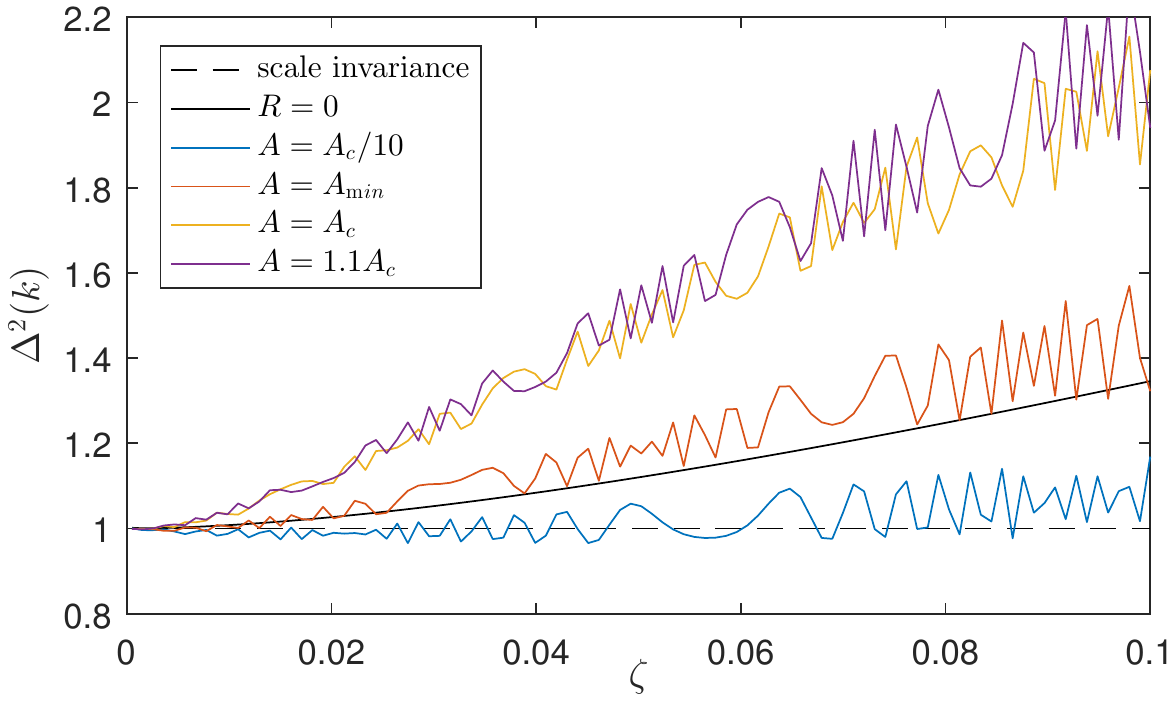}
\caption{$\Delta^2(k)=k^2P(k)$ as a function of in-plane momentum 
$\zeta$, for 2.5 e-folds. Black dashed line represent SIPS. The black solid line corresponds to contact interaction, $R = 0$  ($A = A_c/10$).
The other lines correspond to DDI domination ($R = \sqrt{\pi/2}$), with values of $A$ as specified in the inset. In the long-wavelength limit, 
they all  converge to SIPS. 
The slope of the $R=0$ curve  decreases for increasing number of e-folds, asymptotically 
yieldding SIPS for pure contact interactions.}
\label{fig:20151216_2}
\end{figure}

{From \eqref{eqn:20150926_5}, we have $4H^2 =\omega^2(0)$ (for $\omega_0^2 \ll 4H^2$).  
Setting $\kappa_0 = 50$, $\omega_{z,0} = 2\pi\times 2921$Hz results in 
$H = 183.5\,$sec$^{-1}$. 
Given $n_f $ e-folds of the scale factor $a(t_f)=\exp[n_f]$, the final lab time is
$t_f=(\exp[4 n_f]-1)/4H$.
For 2.5 $e$-folds, then, 
$t_f\sim 30$\,sec in lab time (for $w_{z,0} = 2\pi\times 3952$ Hz, $H = 248.3$\,sec$^{-1}$,
and 2 e-folds, $t_f \sim 3$\,sec)}.
We introduce a momentum cutoff $\zeta_c \lesssim 0.1$ which meets \eqref{eqn:20160411_2} at late times.
Fig.\,\ref{fig:20151216_2} displays $\Delta^2(k)$, 
and clearly shows the deviations from SIPS, occurring for strongly  dipolar interactions.
When $R= 0$, Eq.\,\eqref{eqn:20160411_5} becomes identical to the wave equation in analogue curved spacetime \eqref{eqn:20150926_6}, and SIPS for long wavelengths obtains, cf. Ref.\,\cite{1367-2630-7-1-248} and Fig.\,\ref{fig:20151216_2}.
For high momenta, there is a slight upturn in the spectrum line.
As we increase the number of $e$-folds, this deviation converges to zero; for small wavelengths, it takes longer time to exit the Hubble horizon, and settling down 
requires longer.
Using the power spectrum $\Delta^2(k)$, one again constructs Gaussian random fields, 
and the coordinate-space realization of Fig.\,\ref{fig:GRF_realization} \,(b)--(f) is obtained, 
demonstrating the violation of SIPS for increasing DDI by introducing short-range
correlations.  


Whether SIPS is robust to trans-Planckian physics was studied in \cite{Niemeyer} (also cf.\,\cite{Starobinsky2001}), where scale separation and adiabaticity {in conformal time} 
were established as sufficient conditions for SIPS.
Scale separation reads 
${H}/{c_0} \ll k_{\Pl}(\eta_k)/ a(\eta_k)$,
while adiabaticity holds when 
$
\left| {c_0\ptl_\eta \omega_\eta}/{\omega_\eta^2} \right| \ll 1 \,\,\forall\,\, 
\eta_i < \eta <  \eta_f,
$
where $\omega_\eta(\eta)$ is an effective comoving frame mode frequency 
\cite{Niemeyer,suppl}.  
Furthermore, $\eta_f$ is the `nonadiabatic time' lying between $\eta_i$, the onset of inflation, and $\eta_k$, the horizon crossing, which satisfies 
$
{H}/{c_0} \ll {k}/{a(\eta_f)} \ll {k_\Pl(\eta_k)}/{a(\eta_k)}.
$
Roughly speaking, $\eta_f$ is the moment when the mode stops to behave WKB-like.
For de Sitter spacetime, $a = -c_0 / H\eta$, scale separation  
holds when $k \ll k_{\Pl} (\eta_k)$. 
A given $k$ thus must lie in the linear dispersion regime at 
horizon crossing $k\eta_k = -1$, which is is equivalent to imposing \eqref{eqn:20160411_2} at this point.
Therefore, in our numerical implementation of the Bogoliubov equations which employs \eqref{eqn:20160411_2}, 
scale separation is satisfied automatically.
According to \cite{Niemeyer}, scale separation  usually implies adiabaticity, 
resulting in the robustness of the predictions of the inflationary scenario.
{However, when the spectrum has {(even if only initially)} 
a deep minimum, 
as here, 
adiabaticity can be violated even when scale separation holds, and SIPS breaks down.}

In conclusion, we have found that for contact interactions, $R = 0$, SIPS is retained (in the limit of
many e-folds), while there appear strong deviations from scale invariance in the presence of strong  DDI (Fig.\,\ref{fig:20151216_2}) due to an initially present roton minimum. 
Importantly, 
the influence of the trans-Planckian nonlinear dispersion is manifest even far from criticality at $A_c$.
When a negative slope in the excitation spectrum occurs ($A > A_{\mathrm{min}}$ in \,Fig.\ref{fig:Bogol_Spec}), the power spectrum shows a general tendency of increase at high momenta.
On the other hand, for monotonously increasing spectrum, when $A \leq A_{\mathrm{min}}$, the power spectrum oscillates around the SIPS prediction. 

We stress that the presence of a minimum in the spectrum does {\em not necessarily} 
imply violations of scale invariance. It is possible to construct 
an analytic solution to the full Bogoliubov equations for a spectrum with minimum, which 
displays SIPS \cite{suppl}. The proposed experiment (or variants thereof, possibly with 
other engineered interaction potentials) 
can thus potentially lead to conclusions about the 
trans-Planckian physics of quantum fields in early cosmological stages.
We also note in this regard that SIPS is a kinematical effect for quantum fields in de Sitter spacetime,   
in analogy to Hawking radiation 
from black holes \cite{Visser1998}, and therefore, like the latter, does not 
require the Einstein equations to hold.


Going beyond 
mean-field theory, future perspectives include to study the influence of strong quantum fluctuations of 
high density electrically dipolar gases \cite{Baranov}, prevailing in 
an early, possibly pre-metric 
stage, onto the analogue cosmological evolution 
in the inflationary scenario.


This research was supported by the NRF Korea, Grant No. 2014R1A2A2A01006535. 
\bibliography{tp12}

\begin{thebibliography}{70}
\bibitem [{ren()}]{renorm}%
  \BibitemOpen
  \href@noop {} {\ }\bibinfo {note} {\!\!We neglect the interaction-dependent
  renormalization of $d_z$ in the crossover regime from quasi-2D to 3D
  \cite{Fischer}, which decreases in time as the gas becomes more
  dilute.}\BibitemShut {Stop}%
\end{thebibliography}

\vspace*{80em} 
\newpage

\begin{widetext}
\setcounter{equation}{0}
\setcounter{figure}{0}
\setcounter{table}{0}
\setcounter{page}{1}
\renewcommand{\theequation}{S\arabic{equation}}
\renewcommand{\thefigure}{S\arabic{figure}}

\section{Supplemental Material}

\subsection{Action of the system \label{sec:20160808_1}}

\subsubsection{Dimensional reduction \label{subsec:20160808_2}}

In the limit of zero-point energy of the axial harmonic oscillator greatly exceeding the chemical potential, and for large aspect ratio, the longitudinal and transversal degrees of freedom decouple 
and we can factorize the order parameter $\Psi(\bR,t)$ as follows 
\begin{equation} \label{eqn:20150924_3}
\Psi(\bR,t) = \Psi_r(\br,t)\Phi_z(z)e^{-i\omega_z t/2}.
\end{equation}
Here $\Phi_z(z)$ describes the zero point oscillations in a harmonic oscillator potential, and is given by
$$
\Phi_z(z) = \frac{1}{(\pi d_z^2)^{1/4}}\exp\Bigl[ -\frac{z^2}{2d_z^2}\Bigr],
$$
where $d_z = \sqrt{\hbar/m\omega_z}$ is the oscillator length.
Improved estimates for $d_z$ can be found by treating $d_z$ as a parameter minimizing the Gross-Pitaevski\v\i~ ground-state energy \cite{Fischer, 10.1140/epjd/e2004-00146-7}.

Substituting \eqref{eqn:20150924_3} into the action \eqref{eqn:20150924_1}, and integrating out the $z$ dependence, we obtain the reduced Lagrangian for the horizontal in-plane mode:
\begin{align*}
\cL_r & = \frac{i\hbar}{2}(\Psi_r^*\partial_t {\Psi}_r - \partial_t{\Psi}_r^*\Psi_r) 
    - \frac{\hbar^2}{2m}|\nabla_r \Psi_r|^2 - \frac{m}{2}\omega^2r^2|\Psi_r|^2 - \frac{g_c^{\td}}{2}|\Psi_r|^4 
    - \frac{1}{2} |\Psi_r|^2 \int d^2\br'\, V_{dd}^{\td}(\br - \br')|\Psi_r(\br')|^2,
\end{align*}
where $\nabla_r = (\ptl_x,\ptl_y)$.
The contact coupling is reduced to $g_c^{\td} = g_c/\sqrt{2\pi}d_z$, and the reduced lower-dimensional DDI is given by ($\rho_z = |\Phi_z|^2$)
\begin{equation} \label{eqn:20150924_4}
V_{dd}^{\td}(\br - \br') = \int dzdz'\, V_{dd}(\bR - \bR')\rho_z(z)\rho_z(z'). 
\end{equation} 
We assume the dipoles to be polarized along $z$-direction by an external field, so that their interaction is 
\begin{equation} \label{eqn:20150924_2}
V_{dd}(\bR,t) = \frac{3g_d}{4\pi}\frac{1 - 3z^2/|\bR|^2}{|\bR|^3},
\end{equation}
where $g_d = \mu d_m^2/3$ for magnetic and $g_d = d_e^2/3\epsilon$ for electric dipoles.

The nature of this interaction can be seen clearly by looking at its Fourier space representation.
The Fourier transform of the DDI \eqref{eqn:20150924_2} takes the well-known form 
$$
{V}_{dd}(\bQ) = g_d\Bigl(\frac{3q_z^2}{Q^2} - 1 \Bigr),
$$
where $\bQ = (\bq, q_z)$.
Here and below, we will use asymmetric Fourier convention in which the inverse transform incorporates the whole prefactor.
The Fourier transform of the density profile in $z$-direction $\rho_z$, for homogeneous density in the 2D plane,  
is given by ${\rho}_z(\bQ) = V\delta^{(2)}_{\bq,0}\exp[-q_z^2d_z^2/4]$, where $V$ is the area of the plane. Substituting the inverse Fourier transforms of these expressions into \eqref{eqn:20150924_4}, we obtain \cite{Fischer}
\begin{equation} \label{eqn:20150924_5}
V_{dd}^{\td}(\br - \br') = \frac{g_d}{V}\sum_{\bq} e^{i\bq\cdot(\br -\br')}\biggl\{\frac{2}{\sqrt{2\pi}d_z} - \frac{3q}{2}w\biggl[\frac{qd_z}{\sqrt{2}}\biggr]\biggr\}.
\end{equation}
Here we made use of an integral representation of the error function \cite{abr65}
\begin{equation} \label{eqn:20150924_6}
w(z) := e^{z^2}\mathrm{erfc}(z) = \frac{2z}{\pi}\int_0^\infty \frac{e^{-t^2}}{z^2 + t^2}dt \quad (z>0), 
\end{equation}
where the complementary error function is defined as $\mathrm{erfc}(z) = 1 - \mathrm{erf}(z) = 1 - (2/\sqrt{\pi})\int_0^z \exp(-t^2)dt$.

From \eqref{eqn:20150924_5}, we see that the DDI contributes to the delta-function-like interaction as well as to the nonlocal one.
It is convenient to decompose the total (contact and dipolar) interaction into a sum of effective contact interaction and nonlocal interaction:
\begin{equation} \label{eqn:20150924_7}
g^{\eff}\delta^{(2)}(\br - \br') + U^{\td}(\br - \br'),
\end{equation}
where the effective contact coupling strength is defined by
$$
g^{\eff} := g_c^{\td} + \frac{2g_d}{\sqrt{2\pi} d_z} = \frac{1}{\sqrt{2\pi}d_z}(g_c + 2g_d),
$$
and the nonlocal interaction is written as
$$
U^{\td}(\br - \br') = -\frac{3}{2}\frac{g_d}{V}\sum_{\bq} e^{i\bq\cdot(\br - \br')}qw\biggl[ \frac{qd_z}{\sqrt{2}} \biggr].
$$

\subsubsection{Scaling transformation \label{subsec:20160808_3}}

Having performed dimensional reduction, we consider the action
\begin{equation} \label{eqn:20150924_8}
\begin{aligned}
\cL & = \frac{i\hbar}{2}(\Psi^*\dot{\Psi} - \dot{\Psi}^*\Psi) 
    - \frac{\hbar^2}{2m}|\nabla\Psi|^2 - \frac{m}{2}\omega^2r^2|\Psi|^2 - \frac{g^{\eff}}{2}|\Psi|^4 
    -\frac{1}{2}|\Psi|^2\int d^2\br'\, U^{\td}(\br - \br')|\Psi(\br')|^2,
\end{aligned}
\end{equation}
where we dropped subscripts for conciseness.
We can prescribe an external time dependences not only with a temporal profile of the 
trap frequencies but also to 
$g_c=4 \pi \hbar^2 a_s / m$ and $g_d$ by changing the $s$-wave scattering length $a_s$ via Feshbach resonances \cite{PhysRevLett.81.69, Inouye1998Observation} and using a rotating polarizing field to change $g_d$ \cite{PhysRevLett.89.130401}, respectively.
As a result, the gas cloud will adapt to these changes and will either expand or contract.

A part of this background motion can be accounted for by transforming to a new coordinate system
\begin{equation} \label{eqn:20151001_4}
\bx := \frac{\br}{b(t)},\qquad \tau := \int_0^t  \frac{1}{b^2(t')}dt',
\end{equation}
with a scale factor $b(t)$.
Following \cite{PhysRevA.54.R1753}, 
define $\psi(\bx,\tau)$ by
\begin{equation} \label{eqn:20150924_9}
\Psi(\br,t) := e^{i\Phi}\frac{\psi(\bx,\tau)}{b},
\end{equation}
where $\Phi(\bx,t) = \frac{1}{2}\frac{m}{\hbar}r^2\frac{\partial_t{b}}{b}$ is chosen so as to describe the bulk velocity $\bv = \dot{\bx} = -\frac{1}{b}\frac{\hbar}{m}\nabla\Phi$, while the phase of $\psi$ will represent the residual velocity potential, which can be regarded as small.
Insertion of this ansatz into the action \eqref{eqn:20150924_8} yields
\begin{align}
\cL & = \frac{i\hbar}{2}(\psi^*\ptl_\tau \psi - \ptl_\tau \psi^* \psi) 
    - \frac{\hbar^2}{2m}|\nabla_x\psi|^2 - \frac{m}{2}x^2\left(\frac{d^2b}{dt^2} {b^3} + \omega^2b^4\right)|\psi|^2 - \frac{g^{\eff}}{2}|\psi|^4 
\nonumber
\\ & \quad
    - \frac{g_d}{2g_{d,0}b}\int d^2\bx'\, U^{\td}_0(\bx - \bx')|\psi(\bx)|^2|\psi(\bx')|^2. \label{eqn:20150925_1}
\end{align}
Note that the measure $dtd^2\br$ gives an additional factor of $b^4$ by the relation $dtd^2\br = b^4d\tau d^2\bx$.
The scaled nonlocal interaction is written as
$$
U^{\td}_0(\bx - \bx') = -\frac{3}{2}\frac{g_{d,0}}{V}\sum_{\bk} e^{i\bk\cdot(\bx - \bx')}kw\bigg[ \frac{kd_{z,0}}{\sqrt{2}}\biggr],
$$
where $g_{d,0}$ and $d_{z,0}$ are initial values.
In order to obtain this expression, we have assumed a scaling condition
\begin{equation} \label{eqn:20150925_2}
d_z(t) = d_{z,0}b(t) \quad \text{ or } \quad \omega_z(t) = \frac{\omega_{z,0}}{b^2(t)}.
\end{equation}
We can combine the remaining time dependences into a single factor $f(t)$ by imposing \eqref{eqn:20150925_3}.
Here $g_{c,0}$ is the initial value and $\omega_0$ will be fixed when a specific solution to the scaling equation is chosen.
Examples of analytic solutions to the scaling equation \eqref{eqn:20150925_3} are presented in \eqref{eqn:20150926_4} and \eqref{eqn:20150926_5}.
Under these scaling conditions, the action \eqref{eqn:20150925_1} becomes
\begin{equation} \label{eqn:20150925_4}
\begin{aligned}
\cL & = \frac{i\hbar}{2}(\psi^*\ptl_\tau \psi - \psi\ptl_\tau\psi^*) - \frac{\hbar^2}{2m} {|\nabla_x\psi|^2} 
    -f^2\biggl[\frac{m}{2}\omega_0^2x^2|\psi|^2 + \frac{g^{\eff}_0}{2}|\psi|^4 
    + \frac{1}{2} \int d^2\bx'\, U^{\td}_0(\bx - \bx') |\psi(\bx)|^2|\psi(\bx')|^2 \biggr].
\end{aligned}
\end{equation}


\subsubsection{Hydrodynamic variables and a background solution}

In terms of the Madelung representation for the scaled order parameter, $\psi = \sqrt{\rho}e^{i\phi}$, the equation of motion \eqref{eqn:20150925_5} can be recast as
\begin{align}
\ptl_\tau \rho & = -\frac{\hbar}{m}(\nabla_x\phi \cdot \nabla_x\rho + \rho\nabla_x^2\phi), \notag \\
-\hbar\ptl_\tau \phi & = -\frac{\hbar^2}{2m\sqrt{\rho}}\nabla_x^2\sqrt{\rho} + \frac{\hbar^2}{2m}(\nabla_x\phi)^2 + f^2\frac{m}{2}\omega_0^2x^2 
               +f^2\int d^2\bx'\, V^{\td}_{\intac,0}(\bx - \bx')\rho(\bx'), \label{eqn:20150925_6}
\end{align}
where $V^{\td}_{\intac,0}(\bx - \bx') = g_0^{\eff}\delta^{(2)}(\bx - \bx') + U_0^{\td}(\bx - \bx')$.
If we linearize the fields around stationary background solutions, $\rho = \rho_0 + \delta \rho,\, \phi = \phi_0 + \delta \phi$, the zeroth order equations are the same as \eqref{eqn:20150925_6} with subscripts $0$ attached to the fields, and the first order equations are the Bogoliubov equations \eqref{eqn:20151102_1}.

\begin{figure}[ht]
\includegraphics[width=0.6\textwidth]{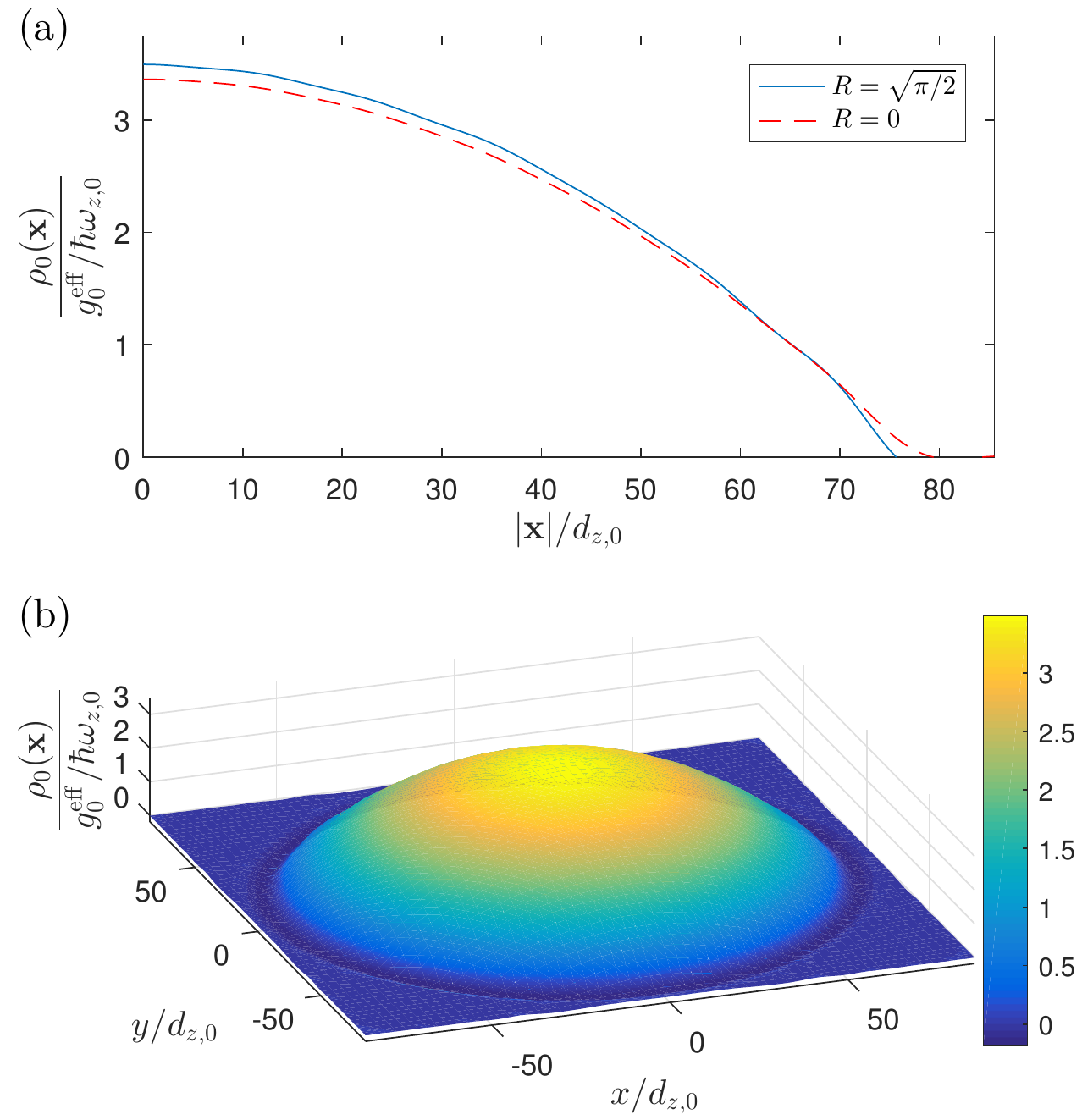}
\caption{The density profile of the gas in units of $g^\eff_0/\hbar\omega_{z,0}$ as a function of radial distance in units of $d_{z,0}$. In (a), blue solid and red dashed line corresponds to DDI-dominant and contact-dominant cases, repectively. We also present a visualization of the gas in (b), using parameters appropriate for 
erbium atoms \cite{Aikawa}. Namely, particle number 
$N = 9.5\times 10^4$, magnetic moment $d_m = 7\,\mu_B$, boson mass $m = 168\,\mathrm{u}$, aspect ratio $\kappa_0 = 30$, and 
transverse trapping frequency $\omega_{z,0} = 2\pi\times 5435$\,Hz. 
}
\label{densityprofile}
\end{figure}

We solve the zeroth order equations assuming vanishingly small residual comoving frame velocity ($\bv_{\rm com} := \frac{\hbar}{m}\nabla_x\phi_0 = 0$) by the ansatz $\psi_0(\bx,\tau) = \sqrt{\rho_0(\bx)}e^{i\phi_0(\tau)}$ \cite{1367-2630-7-1-248}, and neglect the kinetic energy term, which is equivalent to neglecting terms proportional to $\nabla_x^2\sqrt{\rho_0}$.
Then we obtain a spatially constant phase function
\begin{equation} \label{eqn:20150925_7}
\phi_0(\tau) = -\frac{\mu_0}{\hbar} \int_{0}^\tau d\tau'\, f^2(\tau'),
\end{equation}
where $\mu_0$ is initial chemical potential, 
and an integral equation for time independent density profile
\begin{equation} \label{eqn:20150925_8}
\int d^2\bx'\, V^{\td}_{\intac,0}(\bx - \bx')\rho_0(\bx') = \mu_0 - \frac{m}{2}\omega_0^2x^2,
\end{equation}
which can be solved numerically.
Because of the partially attractive nature of DDI, the profile shows enhanced concentration at the center compared to the pure contact case, cf. Fig.\,\ref{densityprofile}.
Also, the anisotropy of the interaction results in the appearance of small wiggles in the density profile 
\cite{PhysRevA.78.041601, PhysRevA.82.023622}.

\subsection{Gravitational analogy and an analytical solution \label{sec:20150928_7}} 

\subsubsection{Effective geometry}

Rewriting the equation \eqref{eqn:20150926_6} in real space, the resulting equation is equivalent to the phases only Lagrangian
\begin{equation} \label{eqn:20150925_19}
\overline{\cL^{(2)}} = \frac{\hbar^2/2}{f^2g_0^{\eff}}(D\delta \phi)^2 - \frac{\hbar^2\rho_0}{2m}(\nabla_x\delta \phi)^2,
\end{equation}
where $D = \ptl_\tau + \bv_{\rm com}\cdot \nabla_x$ is the comoving derivative.

With the metric tensor \eqref{eqn:20150925_20}, the Lagrangian becomes that of a minimally coupled free scalar field in a curved spacetime
\begin{equation} \label{eqn:20150925_21}
\overline{\cL^{(2)}} = \frac{mc_0^2}{2}\sqrt{|g|}g^{\mu\nu} \ptl_\mu \delta\phi \ptl_\nu \delta \phi,
\end{equation}
where repeated indices imply summation over $\mu = 0,\,1,\,2$ and $x^0 = c_0\tau$.

For the background solutions \eqref{eqn:20150925_7} and \eqref{eqn:20150925_8}, $\bv_{\rm com} = \frac{\hbar}{m}\nabla_x\phi_0 = \mathbf{0}$ and $\rho_0$ is time independent.
In this case the line element \eqref{eqn:20150925_20} becomes
\begin{equation} \label{eqn:20150926_1}
ds^2 = \Omega^{-2}\bigl( c_0^2d\tau^2 - a^2d\bx^2 \bigr),\quad a = \frac{1}{f},
\end{equation}
where the conformal factor 
\begin{equation} \label{eqn:20150926_2}
\Omega = \frac{c_0^2m^2}{\hbar^2\rho_0}
\end{equation}
is dimensionless.
Herein, we 
assume that the density 
$\rho_0$ is essentially homogeneous near the center of cloud.
This implies that $\omega_0$, the trapping frequency in the scaled coordinate system, is negligible compared to the time scale of the effective spacetime (that is the Hubble constant $H$, cf. \eqref{eqn:20150926_5}).
Then $\Omega$ becomes just a constant, and the action \eqref{eqn:20150925_21} is invariant under the conformal transformation
\begin{equation} \label{eqn:20160823_3}
\tilde{g}_{\mu\nu} = \Omega^2g_{\mu\nu} \quad \text{ and } \quad \delta \tilde{\phi} = \Omega^{-1/2}\delta \phi,
\end{equation} 
and the resulting metric $\tilde{g}_{\mu\nu}$ assumes the form of FRW universe
\begin{equation} \label{eqn:20150926_3}
ds^2 = c_0^2d\tau^2 - a^2d\bx^2 = \tilde{g}_{\mu\nu}dx^\mu dx^\nu,
\end{equation}
where $\tilde{g}_{\mu\nu} = \mathrm{diag}(1,\,-a^2,\,-a^2)$.
Now we can apply standard techniques of quantum field theory in a FRW universe to obtain independent solutions for $\delta\tilde{\phi}$.
Then the independent solutions for original field $\delta\phi$ will be obtained by $\delta\phi = \Omega^{1/2}\delta\tilde{\phi}$.
In this effective spacetime, the Klein-Gordon (KG) equation for massless, minimally coupled free scalar field,
\begin{equation}
\Box\delta \tilde{\phi} = (1/\sqrt{|\tilde{g}|})\ptl_\mu(\sqrt{|\tilde{g}|}\tilde{g}^{\mu\nu} \ptl_\nu \delta \tilde{\phi}) = 0, 
\end{equation} 
takes the form of \eqref{eqn:20150926_6}.

Here, for ease of connecting the current discussion to a standard cosmological context, we introduce the conformal time
\begin{equation}
\eta := \int_\infty^\tau \frac{c_0}{a(\tau')} \, d\tau',
\end{equation}
which ranges from $-\infty\;(\tau \to -\infty)$ to $0\;(\tau \to \infty)$.
Then the metric \eqref{eqn:20150926_1} takes the conformally flat form $ds^2 = a^2 [ d\eta^2 - d\bx^2]$, and the equation \eqref{eqn:20150926_6} can be recasted in terms of an auxiliary field $\chi_k := \sqrt{a} \delta \tphi_k$ by
\begin{equation} \label{eqn:20160420_1}
\partial^2_\eta \chi_k + \left[ k^2 - \frac{\partial^2_\eta a}{2a} + \frac{(\partial_\eta a)^2}{4a^2} \right] \chi_k = 0 \quad \Leftrightarrow \quad 
\partial^2_\eta \chi_k + \omega^2_\eta (\eta) \chi_k=0. 
\end{equation}
Comparing this equation with Eq.\,(1) in 
\cite{Niemeyer}, one identifies $\omega_\eta$ as an effective comoving frame mode frequency.
The choice of auxiliary field $\chi_k$ is motivated by removing the first derivative term in \eqref{eqn:20150926_6}.

\subsubsection{Mode functions in $2+1$-dimensional de Sitter spacetime} \label{sec:20150928_9}

We consider de Sitter spacetime by setting $a(\tau) = 1/f(\tau) = e^{H\tau}$.
There are several simple analytic solutions to the scaling equation \eqref{eqn:20150925_3} for the realization of analogue de Sitter spacetime.
For example one can consider $b := 1\,\forall\, t$, so that 
scaling time equals lab time, $\tau =t $, and obtain the scale factor evolution 
\begin{equation} \label{eqn:20150926_4}
a^{-2}(t) = e^{-2Ht} = \frac{\omega^2}{\omega_0^2} = \frac{g_c}{g_{c,0}} = \frac{g_d}{g_{d,0}}.
\end{equation}
While this expansion has the advantage of scaling and lab time being identical, $\tau := t$,   
it is experimentally challenging to realize because of the (simultaneously) 
exponentially in  time varying coupling constants.
Another analytic solution, which is found by imposing $g_d := g_{d,0}$, gives \eqref{eqn:20150926_5}.
This solution implies $g_c$ as well as $g_d$ to be constant.
Our numerical analysis is based on this solution.
Note that we assume $\omega_0$ to be negligible compared to $H$ in the quasihomogeneous limit.

A simple parameter can help us understand the underlying physical process and characterize appropriate asymptotic regimes. Define
\begin{equation} \label{eqn:20150926_10}
s := \frac{c_0/H}{a/k} = \frac{c_0k}{Ha},
\end{equation}
which is the ratio of Hubble radius to the physical wavelength of a chosen mode.
The parameter $s$ starts from infinity and finally approaches 0.
Note that, in the de Sitter analogue $a = e^{H\tau}$, the conformal time becomes $\eta = -c_0/Ha$, and the parameter $s$ can be written employing conformal time simply as $s = -k\eta$.
One can see that the horizon crossing time $\eta_k$ of a chosen wavenumber $k$ is determined by $s = 1$ or $k = a(\eta_k) H/c_0$.
In the de Sitter analogue, $a = -c_0/H\eta$, the horizon crossing time is the moment when
\begin{equation} \label{eqn:20160526_1}
k \eta_k = -1.
\end{equation}

The equation \eqref{eqn:20150926_6} can now be written as
\begin{equation} \label{eqn:20160411_4}
\delta \tilde{\phi}_k'' - \frac{1}{s} \delta \tilde{\phi}_k' + \delta \tilde{\phi}_k = 0,
\end{equation}
where prime denotes taking derivative with respect to $s$.

Large $s$ implies that the mode is well inside the Hubble radius and does not feel the curvature of the analogue spacetime.
When $a$ is small, i.e., before the inflation, the condition $s\gg 1$ is satisfied for wide range of $k$ and so all the relevant modes are well inside the Hubble radius.
At this epoch, the second term in \eqref{eqn:20150926_6} can be neglected and we get the WKB solution for time varying frequency $\omega_k = c_0k/a$:
\begin{equation} \label{eqn:20150926_12}
\begin{aligned}
\delta\tilde{\phi}_k \longrightarrow 
\sqrt{\frac{\hbar V}{2ma^2Hs}}\exp(is)
                                     = \sqrt{\frac{\hbar V}{2ma^2\omega_k}}\exp\left( -i\int_\infty^\tau \omega_k(\tau') d\tau' \right) ,
\end{aligned}
\end{equation}
where coefficients are chosen by imposing the normalization condition $\bigl( \delta \tilde{\phi}_ke^{i\bk\cdot\bx}/V,\, \delta \tilde{\phi}_ke^{i\bk'\cdot \bx}/V \bigr)_{\KG} = \delta^{(2)}_{\bk,\bk'}$ and the conserved Klein-Gordon (KG) inner product is defined by \cite{PhysRevA.79.043616, PhysRevD.82.044042}
\begin{equation} \label{eqn:20150926_7}
\begin{aligned}
(f,g)_{\KG} & = i\frac{mc_0}{\hbar} \int d^2\bx \, \sqrt{|\gamma|}f^*(\bx,\tau) \overleftrightarrow{\ptl_n} g(\bx,\tau) \\
            & = i\frac{mc_0^2}{\hbar} \int d^2\bx \, \frac{a^2}{c_0^2} f^*(\bx,\tau) \overleftrightarrow{\ptl_\tau} g(\bx,\tau).
\end{aligned}
\end{equation}
Here, $\gamma$ is the determinant of the metric in the spatial slice $\tau = \mathrm{const.}$, $n^\mu$ is its normal, and $\ptl_n = n^\mu\ptl_\mu$.

We note that, with this choice of coefficients, the canonical commutation relation, $\bigl[\delta \hat{\tilde{\phi}}(\bx,\tau),\, \delta \hat{\tilde{\pi}}(\by,\tau)\bigr] = i\hbar \delta^{(2)}(\bx - \by)$, with conjugate momentum $\delta\tilde{\pi} = \ptl\overline{\cL^{(2)}}/\ptl(\ptl_\tau\delta\tilde{\phi}) = ma^2\ptl_\tau\delta\tilde{\phi}$ holds, and the proper (diagonalized) expression for the energy $\overline{H^{(2)}} = \sum_{\bk} \hbar \omega_k (\ha_\bk^\dag \ha_\bk + 1/2)$ can be obtained.

It is possible to obtain an analytic solution to \eqref{eqn:20160411_4} over the whole range of time.
Following \cite{mukhanov2007introduction,parker2009quantum}, we define a function $F$ by
$$
F(s) = \frac{1}{s} \delta \tilde{\phi}_k.
$$
Then \eqref{eqn:20160411_4} becomes the Bessel equation of order 1:
$$
s^2 F'' + sF' + (s^2 - 1)F = 0,
$$
whose general solution can be written as a linear combination of Bessel functions $J_1$ and $Y_1$ \cite{abr65}.
Thus we obtain
\begin{equation} \label{eqn:20150926_11}
\delta\tilde{\phi}_k(s) = s\Bigl[ A(k)J_1(s) + B(k)Y_1(s)\Bigr].
\end{equation}

We can determine the coefficients $A(k)$ and $B(k)$ by matching this solution with the WKB solution \eqref{eqn:20150926_12} in the $s \to \infty$ limit.
Recalling the asymptotic behavior of Bessel functions \cite{abr65}, we see that, for fixed $\eta$,
$$
B(k) \to i A(k),\quad A(k) \to \sqrt{\frac{\pi\hbar VH}{4mc_0^2k^2}}\quad \text{ as } k\to \infty, 
$$
must be fulfilled in order to match the WKB solution \eqref{eqn:20150926_12} up to a constant phase.

We invoke de Sitter invariance to determine $A(k)$ and $B(k)$ for all $k$.
We observe that the metric \eqref{eqn:20150926_3} with $a(\tau) = e^{H\tau}$ is invariant under the transformation
\begin{align*}
\tau & \to \tau' = \tau + \tau_1,\\
\bx & \to \bx' = e^{-H\tau_1}\bx,
\end{align*}
where $\tau_1$ is arbitrary.
If we define $\bk' := \bk e^{H\tau_1}$, we have $\bk'/a(\tau') = \bk/a(\tau)$ and $\bk\cdot \bx = \bk'\cdot \bx'$.
Thus we obtain
$$
\delta\tilde{\phi}_{k'}(s) = s\Bigl[ A(k')J_1(s) + B(k')Y_1(s)\Bigr],
$$
since $s$ is unchanged when $\tau \to \tau'$ and $k\to k'$.
From the invariance of the metric, it follows that $\delta\tilde{\phi}_{k'}(s)e^{i\bk'\cdot\bx'}/V' = \delta \tilde{\phi}_k(s)e^{i\bk\cdot\bx}/V$ and so
$$
\frac{A(k)}{V} = \frac{A(k')}{V'}
$$
for any $k$ and any $\tau_1$.
Taking $\tau_1 \to \infty$, the r.h.s. converges to $(1/V)\sqrt{\pi\hbar VH/4mc_0^2k^2}$.
Therefore we conclude that $A(k) = \sqrt{\pi\hbar VH/4mc_0^2k^2}$ for any $k$ and the mode function is written as in \eqref{eqn:20150928_10}.

We finally obtain the mode expansion for the phase fluctuation field:
\begin{equation} \label{eqn:20150926_13}
\delta \hat{\tilde{\phi}}(\bx,\tau) = \sum_{\bk} \hta_\bk f^{(0)}_\bk(\bx,\tau) + \hta_\bk^\dag f^{(0)*}_\bk(\bx,\tau),
\end{equation}
where $\hta_\bk$ and $\hta_\bk^\dag$ are time independent creation/annihilation operators obeying the commutation relations $[\hta_\bk, \, \hta_{\bk'}^\dag] = \delta^{(2)}_{\bk,\bk'}$. The mode function is written as
$$
f^{(0)}_\bk(\bx,\tau) = \frac{1}{V}h_k(s)e^{i\bk\cdot \bx}.
$$
The vacuum corresponding to the basis 
$\hta_\bk$ is the Bunch-Davies vacuum \cite{bunch1978quantum}.
Note that at this stage the relation between what $\hta_\bk^\dag$ creates and the Bogoliubov quasiparticles is not clear.
{We will establish a direct connection between them below in \eqref{eqn:20160823_1}.}

\subsubsection{Correlation function}

The correlations of a fluctuating quantum field 
are a measurable quantity in an experimental setup.
Thus we investigate the spatially Fourier-transformed two-point correlation function, which is defined by \cite{Schuetzhold}
$$
C_{\delta\tilde{\phi}}(\bk,\tau) = \int_V d^2\bx\, e^{-i\bk\cdot \bx} \bigl\langle \delta \htphi(0,\tau)\delta \htphi(\bx,\tau)\bigr\rangle.
$$
Now we insert the the mode expansion \eqref{eqn:20150926_13}.
Recalling the asymptotic behavior of the Bessel functions, we have
$$
h_k(s) \to -i\sqrt{\frac{\hbar VH}{\pi m c_0^2}}\frac{1}{k}, \quad \text{ as } s \to 0,
$$
and obtain
\begin{equation} \label{eqn:20150928_2}
C_{\delta\tilde{\phi}}(\bk,\tau) = \frac{|h_k|^2}{V} \:\to\: \frac{\hbar H}{\pi m c_0^2}\frac{1}{k^2}.
\end{equation}
Note that the mode function and correlation function become time independent at late times.
{Thus the density fluctuations determined by \eqref{eqn:20151102_1} vanish at zeroth order}.
In order to obtain nontrivial density fluctuations, one has to take the time dependence of the phase fluctuations into account, which is beyond the zeroth-order frozen part.

\subsubsection{Scale invariant power spectrum \label{subsec:20160411_6}}

The amplitude of quantum fluctuations is always well defined irrespective of whether the particle interpretation of a given field is available \cite{mukhanov2007introduction}. 
One way to characterize the typical fluctuations on scales $L$ is to calculate the variance $\delta \chi_L^2(\tau) = \langle 0 | [\hat{\chi}_L(\tau)]^2|0\rangle$ of the field operator averaged over a region of size $L$:
$$
\hat{\chi}_L(\tau) := \int d^2\bx\, \delta \htphi(\bx, \tau) W_L(\bx),
$$
where $W_L(\bx)$ is a window function which is of order 1 for $|\bx| \lesssim L$ and rapidly decays for $|\bx| \gg L$. It is prototypically specified in terms of Gaussian function $W_L(\bx) = (1/2\pi L^2)\exp(-|\bx|^2/2L^2)$.
Given the mode expansion \eqref{eqn:20150926_13}, after straightforward algebra with an approximation to the Fourier transform of the unit ($L = 1$) window function, $w(\bk) \simeq 2\pi[1 - \theta(k - 1)]$, one can find
$$
\delta\chi_L^2(\tau) \simeq \int_0^{L^{-1}} \frac{dk}{k} \frac{k^2|h_k|^2}{V}.
$$
We define the (two-dimensional version of) power spectrum $P(k)$ to be proportional to the variance per $\ln k$:
$$
k^2P(k) := \Delta^2(k) = \frac{d\delta \chi_L^2}{d\ln k} = \frac{k^2|h_k|^2}{V}, \quad k = L^{-1}.
$$

Another characterization of the power spectrum is as the Fourier transform of the correlation function \cite{peacock1999cosmological}:
\begin{align}
\xi(\bx - \by) & = \langle 0| \delta \htphi(\bx,\tau)\delta \htphi(\by,\tau)| 0 \rangle \notag \\
               & := \frac{1}{V} \sum_\bk P(k) e^{i\bk\cdot(\bx - \by)}. \label{eqn:20150928_1}
\end{align}
from which we have $P(k) = \langle \delta \htphi_\bk \delta \htphi_\bk^\dag \rangle/V = |h_k|^2/V$, where $\delta\htphi_\bk$ is the Fourier transform of the mode expansion \eqref{eqn:20150926_13}.
Note that $P(k)$ is nothing but the correlation function obtained in \eqref{eqn:20150928_2}.
At late times, $\eta \to 0$, the power spectrum $P(k)$ converges to $\hbar H/\pi m c_0^2 k^2$ and we see that $\Delta^2(k) =  k^2P(k)$
becomes independent of $k$.
We thus obtain, after the freezing process, a spectrum in which $\Delta^2(k)$, the variance per $\ln k$ 
\cite{peacock1999cosmological}, is constant. This is called a scale-invariant power spectrum (SIPS): The universe has the same degree of `wrinkliness' on each resolution scale.
One can also understand this concept by observing that $|\delta\tphi_k|^2 \propto \frac{1}{k^2}$, namely the probability amplitude of a fluctuation having wavelength $1/k$ is proportional to the volume(area) of the space that the fluctuation is filling.
Thus the general shape of the fluctuation field will be independent of the resolution scale.

It is commonly argued that the prediction of scale invariance arises because de Sitter space is invariant under time translation: there is no natural origin of time under exponential expansion \cite{peacock1999cosmological}.
At a given moment of time, the only length scale in the model is the horizon size $c_0/H$, so it is inevitable that the fluctuations that exist on this scale are the same at all time.
If one ignores their evolution while they are outside the horizon, the resulting fluctuations give us the scale-invariant or Harrison-Zel'dovich-Peebles spectrum \cite{PhysRevD.1.2726, Zeldovich01101972, Peebles:1970ag}.

Regarding the phase fluctuation field $\delta \htphi(\bx,\tau)$ as a homogeneous and isotropic Gaussian random field \cite{mukhanov2005physical}, i.e. a field whose Fourier coefficients $\delta \tphi_\bk = a_\bk + ib_\bk$ are random variables with probability function of the form,
\begin{equation} \label{eqn:20150928_3}
p(a_\bk,b_\bk) = \frac{1}{\pi\sigma_k^2}e^{-a_\bk^2/\sigma_k^2}e^{-b_\bk^2/\sigma_k^2},
\end{equation}
the correlation function can be expressed as
\begin{align}
\xi(\bx -  \by) & = \frac{1}{V^2} \sum_{\bk,\bk'} \langle \delta \tphi_\bk(\tau) \delta \tphi_{\bk'}^*(\tau)\rangle e^{i\bk\cdot \bx - i\bk'\cdot \by} \notag \\
               & = \frac{1}{V} \sum_{\bk} \sigma_k^2 e^{i\bk\cdot(\bx - \by)}, \label{eqn:20150928_4}
\end{align}
where one has to take into account that $a_{-\bk} = a_\bk$ and $b_{-\bk} = -b_\bk$.
From \eqref{eqn:20150928_3}, one can see that $a_\bk$ and $b_\bk$ are real random variables with standard deviation $\sigma_k/\sqrt{2}$.

Comparing \eqref{eqn:20150928_1} and \eqref{eqn:20150928_4}, we see that the variance of the random variable $\delta \tphi_\bk$ is given by $\sigma_k^2 = P(k) = |h_k|^2/V \to \hbar H/\pi m c_0^2 k^2$, from which we can obtain a real-space realization of the phase fluctuation field $\delta \htphi(\bx,\tau)$ 
as in Fig.\,\ref{fig:GRF_realization}\,(a).

\subsection{Incorporating trans-Planckian deformation} \label{sec:20150930_6}

\subsubsection{Generalized Klein-Gordon equation}

If we rewrite \eqref{eqn:20151102_1} in real space, we obtain
\begin{equation} \label{eqn:20151102_3}
\begin{aligned}
\ptl_\tau \delta \rho & = -\frac{\hbar \rho_0}{m} \nabla_x^2 \delta \phi, \qquad 
\ptl_\tau \delta \phi & = -\frac{f^2g_0^{\eff}}{\hbar} \cW \delta \rho,
\end{aligned}
\end{equation}
where $\cW$ is an integral operator defined by
\begin{equation} \label{eqn:20151102_4}
\cW = \int d^2\bx'\, \biggl[ \frac{1}{V} \sum_\bk \cW_k e^{i\bk\cdot(\bx - \bx')} \biggr] \star(\bx'),
\end{equation}
where $\star$ stands for the argument upon which the integral operator acts.
Solving the second equation in \eqref{eqn:20151102_3} for $\delta \rho$ and substituting into the first equation, we obtain
\begin{equation} \label{eqn:20151102_5}
\ptl_\tau \biggl[ \frac{a^2}{c_0^2} \cW^{-1} \ptl_\tau \delta \tilde{\phi} \biggr] = \nabla_x^2 \delta \tilde{\phi},
\end{equation}
where $\delta \tilde{\phi} = \Omega^{-1/2}\delta \phi$ as in \eqref{eqn:20160823_3}.
This is the `generalized' Klein-Gordon equation with the local Lorentz invariance being broken \cite{PhysRevD.82.044042}.

Rewriting \eqref{eqn:20151102_5} in momentum space, or equivalently, solving the second equation in \eqref{eqn:20151102_1} for $\delta\rho_\bk$ and substituting into the first equation yields \eqref{eqn:20151102_6}.

Let us again introduce an auxiliary field and discuss in the cosmological context.
In order to remove the first derivative term of \eqref{eqn:20151102_6}, we define $\chi_k := \sqrt{a/\cW_k}\delta \tphi_k$, and recast \eqref{eqn:20151102_6} as
\begin{equation} \label{eqn:20160420_2}
\partial_\eta^2 \chi_k + \left[ k^2\cW_k - \frac{\partial^2_\eta a}{2a} \left( 1 - \frac{2a\partial_\eta\cW_k}{\cW_k}\right) + \frac{(\partial_\eta a)^2}{4a^2} \left( 1 - \frac{3a^2 (\partial_\eta\cW_k)^2}{\cW_k^2} + \frac{2a^2 \partial^2_\eta \cW_k}{\cW_k} + \frac{4a \partial_\eta \cW_k}{\cW_k} \right) \right] \chi_k = 0. 
\end{equation}
This equation again corresponds to Eq.\,(1) 
of \cite{Niemeyer}, cf. the relativistic limit above 
in \eqref{eqn:20160420_1}, where 
now the effective comoving frame mode frequency $\omega_\eta$ is the 
square root of the expression in the square brackets. 
It is easily observed that \eqref{eqn:20160420_2} converges to \eqref{eqn:20160420_1} when $\cW_k = 1$, i.e. in the long wavelength limit.
Furthermore, \eqref{eqn:20160420_2} becomes \eqref{eqn:20160420_1} except a factor of $\cW_k$ multiplied to $k^2$ when $\cW_k$ is time independent (or $a$ independent).
This case is discussed in the next subsection.

\subsubsection{An exactly solvable case} \label{subsec:20151102_8}

Solving the general equation \eqref{eqn:20151102_6} requires numerical methods.
We show herein that an analytic solutions under an approximation to the interaction potential and 
introducing a momentum cutoff is feasible. 

We replace the Fourier transform of the interaction ${V}_{\intac,0}^{\td}(\zeta)$ by 
\begin{equation} \label{eqn:20150930_1} \arraycolsep=1pt\def\arraystretch{2.2}
\bar{V}^{\td}_{\intac,0}(\zeta) = \left\{ \begin{array}{l@{\qquad}l} (1 - \dfrac{1}{f^2})\dfrac{g_0^{\eff}\zeta^2}{4A} + {V}^{\td}_{\intac,0}(\zeta) & (\zeta \leq \zeta_c), \\
(1 - \dfrac{1}{f^2})\dfrac{g_0^{\eff}\zeta_c^4}{4A\zeta^2} + {V}^{\td}_{\intac,0}(\zeta) & (\zeta > \zeta_c), \end{array} \right.
\end{equation}
where momentum cutoff $\zeta_c$ is set to include a part of trans-Planckian momenta:
\begin{equation} \label{eqn:20151209_2}
\zeta_c = \zeta_{\mathrm{Pl}}\times \alpha,
\end{equation}
where $\alpha \gtrsim 1$ determines the cutoff location and gives a class of spectrum lines that yields scale invariant power spectra (cf. Eq.\eqref{eqn:20151209_1}). 
Note that initially ($f = 1$) the new potential $\bar{V}^{\td}_{\intac,0}$ coincides with the original one ${V}^{\td}_{\intac,0}$ (Fig.\,\ref{fig:spectrm_approx}\,(a)).
As time passes, the excitation spectrum deviates from the true dispersion.
However, the deviation is localized around the cutoff momentum $\zeta_c$, and the dispersion law at low energies is secured $\forall t$.
\begin{figure}[ht]
\centering
\includegraphics[width=8.6cm]{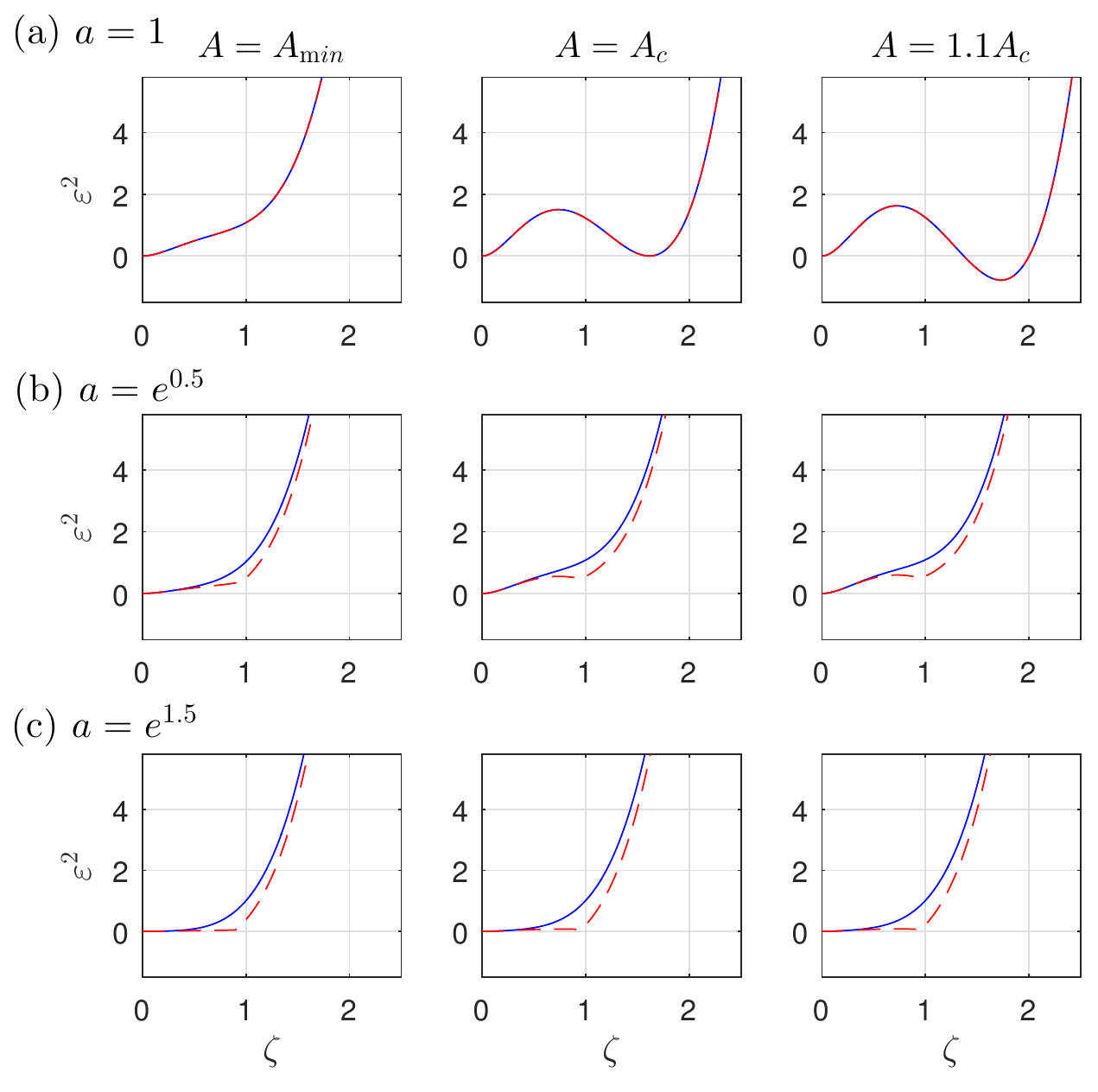}
\caption{The squared excitation spectrum in units of $(\hbar^2/md_{z,0}^2)^2$ at various instants of time. From left to right, the values of $A$ are $A_{\min}$, $A_c$ and $A_c\times 1.1$, repectively. $R = \sqrt{\pi/2}$ for every case.
The cutoff momentum is placed at $\alpha = 1.8$ (cf. \eqref{eqn:20151209_2}). Blue lines represent the original spectrum while the red dashed lines are approximations carried out to obtain an analytic solution. Initially the two coincide and as time evolves they gradually deviate. Note that the deviation is 
however localized around the cutoff momentum $\zeta_c$.}
\label{fig:spectrm_approx}
\end{figure}

Below the cutoff ($\zeta \leq \zeta_c$), the operator \eqref{eqn:20151102_2} becomes time independent
$$
\cW_k = \frac{\zeta^2}{4A} + \frac{1}{g_0^{\eff}}{V}_{\intac,0}^{\td}(\zeta),
$$
and equation \eqref{eqn:20151102_6} becomes
\begin{equation} \label{eqn:20151102_7}
\delta \ddot{\tilde{\phi}}_k + 2H\delta \dot{\tilde{\phi}}_k + \biggl( \frac{c_0 \mathcal{K}(k)}{a} \biggr)^2 \delta \tilde{\phi}_k = 0,
\end{equation}
which is identical to \eqref{eqn:20150926_6} with $k$ being replaced with $\mathcal{K}(k)$ defined by
\begin{equation} \label{eqn:20151102_8}
\mathcal{K}(k)^2 := k^2 \cW_k.
\end{equation}
We can then carry out exactly the same procedure for obtaining the mode functions \eqref{eqn:20150928_10} with $s$ being replaced with $\tilde{s} = c_0\mathcal{K}(k)/Ha$ and with an additional prefactor $\sqrt{\cW_k}$.

With these modified mode functions, the Fourier transformed correlation function, or the power spectrum now becomes (after freezing)
$$
C_{\delta \tphi}(\bk, \tau) = \frac{|h_k|^2}{V} \; \to \; \frac{\hbar H \cW_k}{\pi m c_0^2\mathcal{K}(k)^2},
$$
and the variance per $\ln k$ becomes
\begin{equation} \label{eqn:20151209_1}
\Delta^2(k) = \frac{2\hbar H \cW_k}{m c_0^2} \frac{k^2}{\mathcal{K}(k)^2} = \frac{2\hbar H}{m c_0^2},
\end{equation}
which still is scale invariant.
Therefore, this type of trans-Planckian deformation implied by \eqref{eqn:20150930_1} has no effect on the power spectrum and on the matter distribution after the freezing process and SIPS is retained.

\subsubsection{Numerical implementation of the full Bogoliubov equation}

Let us first analyze the condition \eqref{eqn:20160411_2} in detail.
Fig.\,\ref{fig:20160412_1} shows the plots of $G(\zeta)^2$ and $(a_1\zeta)^2/4A$ for various values of $A$ where $a_1$ is the final value of the scale factor.
We see that, for the validity of the gravitational analogy, one can pose the later time Planck scale to be $\zeta_{\mathrm{Pl}} \lesssim 0.1$.
If $\zeta <0.05$, the condition \eqref{eqn:20160411_2} is safely satisfied for all cases.

\begin{figure}[ht]
\includegraphics[width=8.6cm]{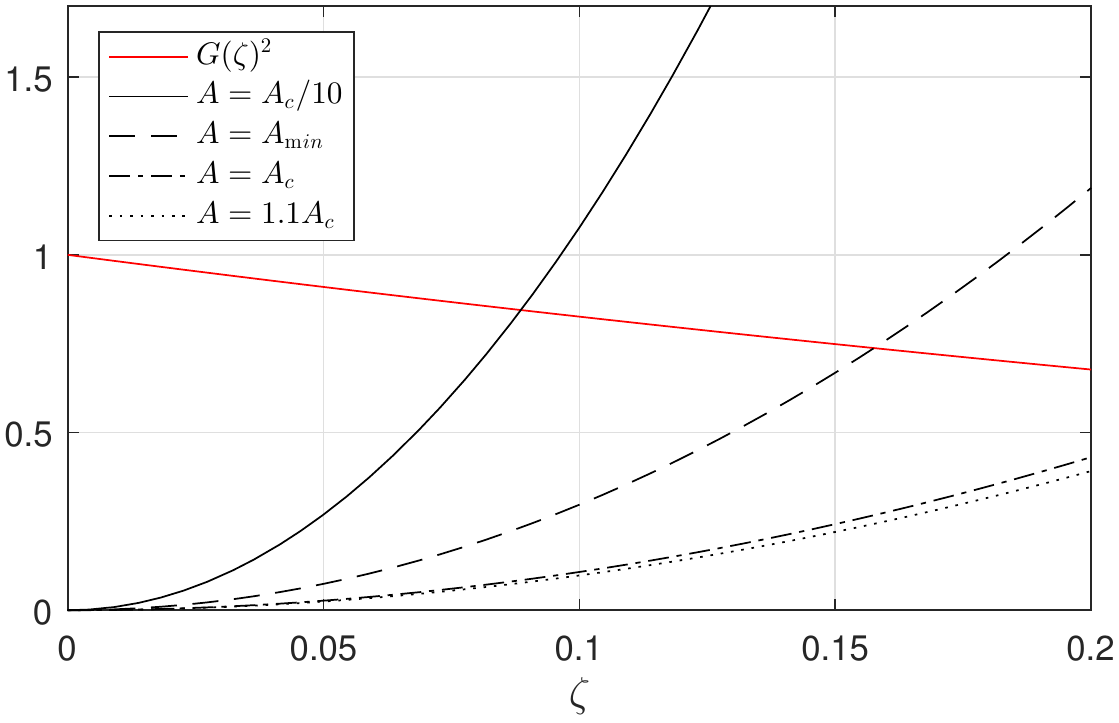}
\caption{Plots of $G(\zeta)^2$ and $(a_1\zeta)^2/4A$ for various values of $A$. Here the final value of the scale factor $a_1$ is assumed to be $e^{5/2}$, i.e., 2.5 $e$-folds of expansion.}
\label{fig:20160412_1}
\end{figure}

Before inflation, $s \to \infty$ and the second term in \eqref{eqn:20160411_5} becomes negligible.
Therefore, one finds that the mode functions would converge to a WKB solution as $a \to 0$:
$$
\delta\tphi_k \to \sqrt{\frac{G(\zeta)\hbar VHs}{2mc_0^2k^2}}\exp(iG(\zeta)s),
$$
where the coefficient is determined by the normalization condition $\bigl( \delta \tilde{\phi}_ke^{i\bk\cdot\bx}/V,\, \delta \tilde{\phi}_ke^{i\bk'\cdot \bx}/V \bigr)_{\mathcal{W}-\KG} = \delta^{(2)}_{\bk,\bk'}$ where the Generalized Klein-Gordon ($\mathcal{W}$-KG) inner product is defined by the equation \eqref{eqn:20151001_2}.
This solution and its derivative provides initial conditions to the second order differential equation \eqref{eqn:20160411_5}.
Final values (after inflation) of the mode functions $h_k$ then give the power spectrum via $P(k) = |h_k|^2/V$ and $\Delta^2(k) = k^2P(k)$.

\subsection{Measurement} \label{sec:20151116_2}

\subsubsection{Bogoliubov transformation to the instantaneous Minkowski vacuum at late times} \label{sec:20151001_3}

Since the de Sitter expansion is not asymptotically flat at late times, a vacuum state cannot be unambiguously defined for late times.
However, the experimental verification obviously requires a choice of Fock vacuum and that choice should lead to physically reasonable results.
We therefore assume, following \cite{PhysRevA.76.033616}, that the expansion stops at some chosen moment of time $\tau_1$ and the gas becomes stationary, in other words, $f(\tau) = e^{-H\tau}$ for $\tau < \tau_1$, and $f(\tau) := f_1$ for $\tau \geq \tau_1$.

Suppose that we have obtained a complete set of ``in'' mode functions $f_\bk^{(0)}$ for $\tau<\tau_1$, 
{e.g. one obtained under \eqref{eqn:20151102_8}}:
\begin{equation} \label{eqn:20150928_11}
f_\bk^{(0)}(\bx,\tau) = \frac{1}{V}h_k(\eta) e^{i\bk\cdot \bx},
\end{equation}
where the temporal part is given by
$$
h_k(\eta) = \sqrt{\frac{\pi\hbar V \cW_k}{4m a^2 H}}\Bigl\{ J_1(-\mathcal{K}(k)\eta) + iY_1(-\mathcal{K}(k)\eta)\Bigr\},
$$
where $\mathcal{K}(k)$ is as defined in \eqref{eqn:20151102_8}.
And suppose that a complete set of ``out'' mode functions $f_\bk^{(1)}$ which defines the vacuum state at late times $\tau > \tau_1$ is given, e.g. one consists of
\begin{equation} \label{eqn:20151102_9}
f_\bk^{(1)}(\bx,\tau) = \frac{1}{V}\sqrt{\frac{\hbar V\cW_k}{2ma_1^2\omega_{k1}}}e^{-i\omega_{k1}(\tau - \tau_1)}e^{i\bk\cdot \bx},
\end{equation}
where $\omega_{k1} := c_0\mathcal{K}(k)/a_1$.
This is a solution to the Bogoliubov equation \eqref{eqn:20151102_7} with $a := a_1$ and $H = \frac{\dot{a}}{a}$ set equal to zero, which  represents the late time behavior of the equation.
The coefficients are fixed by imposing the normalization conditions ($\lambda=0,1$)
\begin{equation} \label{eqn:20151001_1}
\begin{aligned}
(f_\bk^{(\lambda)},f_{\bk'}^{(\lambda)})_{\text{$\mathcal{W}$-$\KG$}} = \delta^{(2)}_{\bk,\bk'},\qquad 
(f_\bk^{(\lambda)},f_{\bk'}^{(\lambda)*})_{\text{$\mathcal{W}$-$\KG$}} = 0,\qquad 
(f_\bk^{(\lambda)*},f_{\bk'}^{(\lambda)*})_{\text{$\mathcal{W}$-$\KG$}} & = -\delta^{(2)}_{\bk,\bk'}, 
\end{aligned}
\end{equation}
where the generalized KG product ($\cW$-KG inner product) is defined by \cite{PhysRevD.82.044042}
\begin{equation} \label{eqn:20151001_2}
(f,g)_{\text{$\cW$-$\KG$}} = i\frac{mc_0^2}{\hbar} \int d^2\bx\, \frac{a^2}{c_0^2} f^*(\bx,\tau) \overleftrightarrow{\cW^{-1}\ptl_\tau}g(\bx,\tau).
\end{equation}
Note that $\cW$-KG inner product converges to the standard relativistic KG product \eqref{eqn:20150926_7} in the limit $\cW \to 1$.

The task at hand is to represent the ``in'' mode functions at $\tau> \tau_1$ as a linear combination of the ``out'' mode functions, i.e. finding the Bogoliubov coefficients $\alpha_k,\, \beta_k$ in the expression
\begin{equation} \label{eqn:20150930_4}
f_\bk^{(0)} = \alpha_k^*f_\bk^{(1)} + \beta_k^*f_{-\bk}^{(1)*}
\end{equation}
for $\tau > \tau_1$.
Then the creation/annihilation operators for ``in'' and ``out'' states will be related by 
$$
\hta_\bk^{(1)} = (f_\bk^{(1)},\,\delta \htphi)_{\text{$\cW$-KG}} = \alpha_k^*\hta_\bk^{(0)} + \beta_k\hta_{-\bk}^{(0)\dag}.
$$
Since $H = {\dot{a}}/{a}$ in \eqref{eqn:20151102_7} changes at $\tau = \tau_1$ in a discontinuous manner, the mode functions and their derivatives must be matched at this point:
\begin{align*}
f_\bk^{(0)}(\tau_1) & = \alpha_k^*f_\bk^{(1)}(\tau_1) + \beta_k^*f_{-\bk}^{(1)*}(\tau_1), \\
\ptl_\tau f_{\bk}^{(0)}(\tau_1) & = \alpha_k^*\ptl_\tau f_{\bk}^{(1)}(\tau_1) + \beta_k^*\ptl_\tau f_{-\bk}^{(1)*}(\tau_1),
\end{align*}
where we suppressed $\bx$ dependence for conciseness.
In the case of \eqref{eqn:20150928_11} and \eqref{eqn:20151102_9}, solving this equation yields
\begin{equation} \label{eqn:20150930_5}
\begin{aligned}
\alpha_k^* & = \sqrt{\frac{\pi\omega_{k1}}{8H}}\biggl\{ J_1 + Y_1' + \frac{HY_1}{\omega_{k1}} + i\biggl[Y_1 - J_1'-\frac{HJ_1}{\omega_{k1}}\biggr]\biggr\}, \\
\beta_k^* & = \sqrt{\frac{\pi\omega_{k1}}{8H}}\biggl\{ J_1 - Y_1'-\frac{HY_1}{\omega_{k1}} + i\biggl[Y_1 + J_1'+\frac{HJ_1}{\omega_{k1}}\biggr]\biggr\},
\end{aligned}
\end{equation}
where the arguments of the Bessel functions are $-\mathcal{K}(k)\eta_1$.
Note that, if the normalization conditions, \eqref{eqn:20151001_1}, are applied to \eqref{eqn:20150930_4}, then one obtains the correct bosonic Bogoliubov unitarity condition $|\alpha_k|^2 - |\beta_k|^2 = 1$.
This relation can also be checked from \eqref{eqn:20150930_5} by direct computation.

If the initial state is assumed to have no excitations, the quantum state is the initial vacuum denoted by $|0\rangle_{(0)}$, i.e., $\hta_{\bk}^{(0)}|0\rangle_{(0)} = 0$.
We consider the Heisenberg picture and the state for $\delta \htphi$ is time independent.
Then the expected number of quasiparticles with momentum $\bk$ after inflation is calculated to be 
\begin{equation} \label{asymptoticN}
_{(0)}\langle 0| \hat{N}_\bk^{(1)} |0\rangle_{(0)} = |\beta_k|^2 \; \to \; \frac{1}{2\pi|\mathcal{K}(k)\eta_1|^3},
\end{equation}
as $\eta_1 \to 0$.

\subsubsection{Translation of cosmological into lab-frame Bogoliubov quasiparticle excitations}

Quantum excitations in BECs can, on the one hand, be analyzed within the Bogoliubov formalism by directly perturbing the Gross Pitaevski\v\i~ equation.
On the other hand, the phase perturbations of the condensate obey a modified Klein-Gordon equation, and a corresponding quantization can be carried out as in \eqref{eqn:20150926_13}.

In order to connect quantum physics in curved spacetime to the behavior of a realistic quantum fluid, Leonhardt \emph{et al.} \cite{1464-4266-5-2-357} investigated the Hawking effect within the Bogoliubov theory of the elementary excitations in BEC.
A more detailed correspondence 
was discussed by Jain \emph{et al.} \cite{PhysRevA.76.033616}, giving an analytical expression for the analogue cosmological particle creation spectrum in terms of the Bogoliubov mode functions in the case of a homogeneous BEC.

Kurita \emph{et al.} \cite{PhysRevA.79.043616} demonstrated the equivalence of the two procedures in the long-wavelength acoustic limit.
They showed that the number of quanta in analogue spacetime is different from that of Bogoliubov quasiparticles, unless the corresponding field is normalized correctly.

Barcel\'o \emph{et al.} \cite{PhysRevD.82.044042} consolidated the equivalence of the two approaches by generalizing the Klein-Gordon formalism beyond the limit of validity of the acoustic approximation.
They showed that both formalism lead to the same concept of positive and negative solutions.
This line of research allows us to establish a deep conceptual connection between the two formalisms, the first one being inherently nonrelativistic while the second is relativistic, up to corrections which are vanishingly small for long wavelengths.

In the following, we discuss the measurement implications of the predictions of previous sections, based on a generalized version of the theory formulated in \cite{PhysRevD.82.044042}.

Under the scaling transformation \eqref{eqn:20150924_9} and the scaling conditions \eqref{eqn:20150925_2} and \eqref{eqn:20150925_3}, the Heisenberg equation of motion for the field operator $\hpsi(\bx,\tau)$ reads
\begin{equation}
i\hbar\ptl_\tau\hpsi = \biggl[ - \frac{\hbar^2}{2m}\nabla_x^2 + f^2 \frac{m}{2}\omega_0^2x^2 
+ f^2 \int d^2\bx' \,V_{\intac,0}^{\td}(\bx - \bx')\hpsi^\dag(\bx')\hpsi(\bx')\biggr]\hpsi.
\end{equation}
Expanding the field operator in canonical way, $\hpsi = \psi_0 + \delta \hpsi$, we obtain the GP equation \eqref{eqn:20150925_5} for the order parameter $\psi_0$, and the Bogoliubov equation \cite{2001camw.book....1C}
\begin{equation} \label{eqn:20150909_2}
i\hbar \ptl_\tau \delta \hpsi = (\mathcal{H} + \mathcal{A})\delta \hpsi + \mathcal{B}\delta \hpsi^\dag,
\end{equation}
where
\begin{align}
\cH & = -\frac{\hbar^2}{2m}\nabla_x^2 + \frac{\hbar^2}{2m\sqrt{\rho_0}}\nabla_x^2\sqrt{\rho_0} - \frac{1}{2}mv_0^2 - \hbar\ptl_\tau \phi_0, \label{eqn:20150909_3}\\
\cA & = f^2\psi_0(\bx)\int d^2\bx'\, V_{\intac,0}^{\td}(\bx - \bx') \psi_0^*(\bx') \star(\bx'), \notag\\
\cB & = f^2\psi_0(\bx)\int d^2\bx'\, V_{\intac,0}^{\td}(\bx - \bx') \psi_0(\bx') \star(\bx'). \notag
\end{align}
In deriving \eqref{eqn:20150909_3}, we have used \eqref{eqn:20150925_6}.
The $\star$ stands for the argument upon which $\cA$ and $\cB$ acts.

Note that Eq.\,\eqref{eqn:20150909_2} is a complex equation and is nonlinear: If $\delta\psi$ is a solution, then $\alpha\delta \psi$ is not unless $\alpha$ is real.
Therefore we cannot directly perform a mode expansion to find the general solution.
In order to overcome this problem, we enlarge the space: We introduce the spinor field
$$
\delta\underline{\Upsilon} = \begin{pmatrix} \delta \psi \\ \delta \bpsi \end{pmatrix},
$$
subject to the evolution equation
\begin{equation} \label{eqn:20150909_6}
i\hbar \ptl_\tau \delta \underline{\Upsilon} = \cM \delta \underline{\Upsilon}, \qquad \cM = \begin{pmatrix} \cH + \cA & \cB \\ -\cB^* & -\cH - \cA^* \end{pmatrix}.
\end{equation}
This equation is now linear, and the solutions to the Bogoliubov equation \eqref{eqn:20150909_2} are obtained by restricting the solutions of \eqref{eqn:20150909_6} by the condition
\begin{equation} \label{eqn:20150909_7}
\delta \bar{\psi} = \delta \psi^*, \quad \text{ or }\quad  \sigma_x\delta \underline{\Upsilon}^* = \delta \underline{\Upsilon},
\end{equation}
where $\sigma_{x,y,z}$ are the Pauli matrices.

We introduce here a conserved ``Bogoliubov'' inner product
$$
\langle \delta \underline{\Upsilon} | \delta \underline{\Upsilon}' \rangle_{\mathrm{B}} = \int d^2\bx' \, \delta \underline{\Upsilon}^\dag \sigma_z\delta \underline{\Upsilon}'.
$$
One can check that the operator $\cM$ is self-adjoint with respect to this inner product
$$
\langle \delta \underline{\Upsilon} | \cM \delta \underline{\Upsilon}'\rangle_{\mathrm{B}} = \langle\cM \delta \underline{\Upsilon}|\delta \underline{\Upsilon}'\rangle_{\rB}.
$$
This implies that the ``Bogoliubov'' inner product is conserved for solutions of \eqref{eqn:20150909_6}.
Note that this inner product is not positive definite, since it satisfies
$$
\langle \sigma_x\delta \underline{\Upsilon}^*|\sigma_x \delta \underline{\Upsilon}^{\prime *} \rangle_{\rB} = -\langle \delta \underline{\Upsilon}' |\delta \underline{\Upsilon}\rangle_{\rB},
$$
and so the physical solutions, i.e. those that satisfy $\sigma_x\delta \underline{\Upsilon}^* = \delta \underline{\Upsilon}$, have zero norm.

The evolution operator $\cM$ is self-adjoint in a non-positive-definite inner product space, and therefore it may have complex eigenvalues.
We will assume that the condensate is stable and $\cM$ has complete othonormal set of eigenspinors with real eigenvalues \cite{PhysRevD.82.044042}.
One can easily check that $\sigma_x\cM\sigma_x = -\cM^*$ holds, and in view of this property, one can see that if 
$$
U_{k} = \begin{pmatrix} u_{\bk} \\ v_{\bk} \end{pmatrix}
$$
is an eigenspinor of $\cM$ with eigenvalue $\omega_{\bk}$, then $V_{\bk}^* = \sigma_x U_{\bk}^*$ is another eigenspinor of $\cM$ with eigenvalue $-\omega_{\bk}$.
Furthermore, the modes $U_{\bk}$ and $V_{\bk}^*$ are orthogonal and can be chosen orthonormal in the Bogoliubov inner product:
$$
\langle U_{\bk} | U_{\mathbf l} \rangle_{\rB} = \delta^{(2)}_{\bk,{\mathbf l}},\quad \langle U_{\bk} |V_{\mathbf l}^*\rangle_{\rB} = 0,\quad \langle V_{\bk}^*|V_{\mathbf l}^*\rangle_{\rB} = -\delta^{(2)}_{\bk,{\mathbf l}}.
$$
Any spinor solution $\delta\underline{\Upsilon}$ of Eq.\,\eqref{eqn:20150909_6} can be expanded in this basis:
$$
\delta \underline{\Upsilon} = \sum_{\bk} b_{\bk} U_{\bk} + c_{\bk}^* V_{\bk}^*.
$$
Note that the modes $U_{\bk}$ and $V_{\bk}^*$ themselves are not physical, while physical solutions are linear combinations of them.

Now the mode expansion for the physical spinor field becomes of the form
$$
\delta \hat{\underline{\Upsilon}} = \sum_{\bk} \hat{\underline b}_{\bk} U_{\bk} + \hat{\underline b}_{\bk}^\dag V_{\bk}^*,
$$
where $\hat{\underline b}_{\bk}$ and $\hat{\underline b}_{\bk}^\dag$ are operators for Bogoliubov quasiparticles.

The (physical or unphysical) spinor field $\delta \underline{\Upsilon}$ corresponds to (complexified) density and phase fluctuations by
\begin{equation} \label{eqn:20150910_11}
\begin{aligned}
\delta \rho & = \sqrt{\rho_0} \bigl( e^{-i\phi_0}\delta \psi + e^{i\phi_0}\delta \bar{\psi} \bigr), \\
\delta \phi & = \frac{1}{2i\sqrt{\rho_0}} \bigl( e^{-i\phi_0}\delta \psi - e^{i\phi_0}\delta \bar{\psi} \bigr).
\end{aligned}
\end{equation}
The condition \eqref{eqn:20150909_7} that $\delta \psi$ and $\delta \bar{\psi}$ represent physical solutions to the Bogoliubov equation \eqref{eqn:20150909_6} translates into reality conditions for $\delta \rho$ and $\delta \phi$.
The density and current operators are then expanded as $\hrho = \hpsi^\dag\hpsi = \rho_0 + \delta \hrho$ and $\hat{\mathbf{j}} = (\hbar/2mi)( \hpsi^\dag \nabla \hpsi - \nabla \hpsi^\dag \hpsi ) = \rho_0\bv_{\rm com} + \bv_{\rm com}\delta\rho + (\rho_0\hbar/m)\nabla_x\delta \hphi$.
In addition, from the bosonic commutation relations $[\delta \hpsi(\bx), \delta \hpsi^\dag(\bx')] = \delta^{(2)}(\bx - \bx')$ etc., one obtains $[\delta \hrho(\bx), \delta \hphi(\bx') ] = i\delta^{(2)}(\bx - \bx')$, i.e., the density and phase fluctuations are canonically conjugate fields.
By the relation \eqref{eqn:20150910_11}, there is one-to-one correspondence between spinor fields $\delta \underline{\Upsilon}$ and complexified density and phase fluctuations $\delta \rho,\; \delta \phi$. {Provided they are physical solutions, $\delta \rho$ and $\delta \phi$ are related by \eqref{eqn:20151102_1}.}

One can readily derive
\begin{equation} \label{eqn:20151001_6}
\langle \delta \underline{\Upsilon} |\delta \underline{\Upsilon}' \rangle_{\rB} = (\delta \tilde{\phi},\delta \tilde{\phi}' )_{\text{$\cW$-KG}},
\end{equation}
where $\delta \tilde{\phi} = \Omega^{-1/2}\delta\phi$, and $\cW$-KG inner product is as defined in \eqref{eqn:20151001_2}.
For a given set of mode functions $\{ f_\bk^{(\lambda)} \}$ for the field $\delta \tilde{\phi}$, 
{which for example were obtained in \eqref{eqn:20150928_11} and \eqref{eqn:20151102_9}}, 
one can find corresponding mode functions $\{ U_\bk^{(\lambda)} \}$ for the spinor field,
and this gives an exact relation between analogue cosmological particles $\hta_\bk^{(\lambda)}$ and Bogoliubov quasiparticles 
$\hat{\underline b}_\bk^{(\lambda)}$:
\begin{equation} \label{eqn:20160823_1}
\hta^{(\lambda)}_\bk = (f_\bk^{(\lambda)},\delta \htphi)_{\text{$\cW$-KG}} = \langle U_{\bk}^{(\lambda)} | \delta \hat{\underline{\Upsilon}}\rangle_{\rB} = \hat{\underline b}_\bk^{(\lambda)}.
\end{equation}
Therefore the number operator of cosmological particles is identical with that of Bogoliubov quasiparticles:
\begin{equation}
\hta_\bk^{(\lambda)\dag}\hta_{\bk}^{(\lambda)} = \hat{\underline b}_\bk^{(\lambda)\dag}\hat{\underline b}_\bk^{(\lambda)}. 
\end{equation}
We note here that the operators $\hta_\bk^{(\lambda)}$ and $\hat{\underline b}_\bk^{(\lambda)}$ correspond to particles that are 
detected in the comoving frame \eqref{eqn:20151001_4}.
However, experiments obviously implement particle detection in the lab frame.
Therefore, one more translation into the lab frame is needed, and is specified below.

\subsubsection{Translation into lab frame variables}

When a normalized mode function $\sqrt{\Omega}f_\bk^{(\lambda)}$ for the field $\delta \phi = \sqrt{\Omega}\delta \tilde{\phi}$ is given, one can get a mode function for the field $\delta \rho$ by the relation
\begin{equation} \label{eqn:20160823_7}
\delta \rho = -\frac{a^2\hbar}{g_0^{\eff}} \cW^{-1} \ptl_\tau \delta \phi,
\end{equation}
which is immediate from the second equation of \eqref{eqn:20151102_3}.
Then one gets the mode functions for $\delta \underline{\Upsilon}$ via
\begin{equation} \label{eqn:20160823_4} \arraycolsep=1pt\def\arraystretch{2.2}
\delta\underline{\Upsilon} = \begin{pmatrix} \delta \psi \\ \delta \bpsi \end{pmatrix} =  \begin{pmatrix} e^{i\phi_0}\biggl[ \frac{1}{2\sqrt{\rho_0}}\delta \rho + i\sqrt{\rho_0}\delta \phi \biggr] \\
                        e^{-i\phi_0}\biggl[ \frac{1}{2\sqrt{\rho_0}}\delta \rho - i \sqrt{\rho_0} \delta \phi \biggr] \end{pmatrix},
\end{equation}
which have already been normalized by \eqref{eqn:20151001_6}.
The perturbed field $\delta \psi$ of the scaled order parameter is related to that of the original Bose field
in the lab frame by ($\Phi=mr^2 \partial_t b/2\hbar b $)
\begin{equation} \label{eqn:20160823_5}
\delta \Psi = \frac{e^{i\Phi}}{b} \delta \psi, \quad \delta \bar{\Psi} = \frac{e^{-i\Phi}}{b} \delta \bar{\psi}.
\end{equation}
The normalization should however still be verified for this field: We form a spinor field
\begin{equation} \label{eqn:20160823_6}
\delta \Upsilon = \begin{pmatrix} \delta \Psi \\ \delta \bar{\Psi} \end{pmatrix} = \frac{1}{b} \begin{pmatrix} e^{i\Phi}\delta \psi \\ e^{-i\Phi} \delta \bar{\psi} \end{pmatrix},
\end{equation}
and introduce the Bogoliubov inner product
\begin{align}
\langle \delta \Upsilon | \delta \Upsilon' \rangle_{\rB} & = \int d^2\br\, \delta \Upsilon^\dag \sigma_z \delta \Upsilon'=\int d^2\bx\, \delta \underline{\Upsilon}^\dag \sigma_z \delta \underline{\Upsilon}'\nonumber\\
              & = \langle \delta \underline{\Upsilon} |\delta \underline{\Upsilon}' \rangle_{\rB} = (\delta \tilde{\phi},\delta \tilde{\phi}' )_{\text{$\cW$-KG}}.
\end{align}
This implies that the cosmological particles are equivalent to the Bogoliubov quasiparticles observed in the lab frame provided the mode functions are chosen consistent with \eqref{eqn:20160823_7}, \eqref{eqn:20160823_4}, \eqref{eqn:20160823_5}, and \eqref{eqn:20160823_6}. It leads to the
lab frame Bogoliubov  quasiparticle operators when expansion stops, see above discussion between Eqs.\,\eqref{eqn:20150928_11} and \eqref{asymptoticN}, being given by  
$\hat b_{\bk/b_1}^{(1)}= \hta_{\bk}^{(1)} $, where
$b_1 := b(t_1)$ is the final scale factor and $\hat b_\bk$ are the annihilation operators associated to $\delta \Upsilon$.

\end{widetext}

\end{document}